\documentclass[aps,11pt,prd,%preprint,
nofootinbib,superscriptaddress]{revtex4-2}

\usepackage{amssymb, amsfonts, amsmath, bm}
\usepackage[%dvipdfmx
]{graphicx}
\usepackage{color}
\usepackage{mathrsfs}
\usepackage[%dvipdfmx
]{hyperref}
\hypersetup{%
 colorlinks=true,
  linkcolor=blue,   
  citecolor=blue,   
  urlcolor=blue
}
\newcommand{\orcid}[1]{%
  \href{https://orcid.org/#1}{%
   \IfFileExists{orcid.pdf}{\includegraphics[height=0.7em]{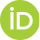}}{}%
 }%
}
\begin{document}
\title{
Strong deflection of massive particles via the geodesic deviation equation
}
\author{Takahisa Igata\:\!\orcid{0000-0002-3344-9045}
}
\email{takahisa.igata@gakushuin.ac.jp}
\affiliation{%
Department of Physics, Faculty of Science, Gakushuin University, Tokyo 171-8588, Japan}
\author{Yohsuke Takamori\:\!\orcid{0000-0002-2298-195X}}
\email{takamori@wakayama-nct.ac.jp}
\affiliation{National Institute of Technology (KOSEN), Wakayama College, Wakayama 644-0023, Japan}
\date{\today}

\begin{abstract}
We develop a formulation of the strong deflection limit for the scattering of particles following timelike geodesics in asymptotically flat, static, and spherically symmetric spacetimes. For fixed specific energy, as the angular momentum approaches its critical value from above, the particle passes arbitrarily close to the associated unstable circular orbit, undergoes many windings around it, and the deflection angle diverges logarithmically. Using the geodesic deviation equation, we show covariantly that the coefficient of this logarithmic divergence is determined by the radial instability exponent of the critical trajectory, defined per unit azimuthal angle. We express this instability exponent in terms of local curvature data on the unstable circular orbit, thereby providing both kinematic and geometric interpretations of the strong deflection limit. In general relativity, its matter dependence enters only through a single local scalar combination constructed from the static-frame energy density and the principal radial and tangential pressures.
\end{abstract}

\maketitle

%%%%%
\section{Introduction}
\label{sec:1}
%%%%%
Radially unstable circular orbits near compact objects underpin a wide class of strong-field observables~\cite{Synge:1966okc,Bardeen:1973}. Event Horizon Telescope images of M87*~\cite{EventHorizonTelescope:2019dse} and Sgr~A*~\cite{EventHorizonTelescope:2022wkp} have highlighted the role of near-critical photon trajectories that whirl around unstable photon orbits (circular in Schwarzschild and spherical in Kerr)~\cite{Luminet:1979,Fukue:1988,Falcke:1999pj,Takahashi:2004xh} before escaping to infinity or plunging inward~\cite{Teo:2003ltt,Igata:2019pgb,Gralla:2019xty}. This behavior motivates analytic control of the strong-deflection regime and naturally leads to the strong-deflection limit (SDL).

In the SDL, scattering trajectories pass arbitrarily close to an unstable circular orbit (the photon sphere for null geodesics), and the deflection angle diverges logarithmically~\cite{Darwin:1959}. The deflection angle is commonly written as $\hat{\alpha}\simeq -\bar{a}\log \varepsilon+\bar{b}$, where $\varepsilon$ measures the distance from criticality (e.g., the fractional deviation of the impact parameter from its critical value, $\varepsilon=\eta/\eta_{\mathrm{c}}-1$)~\cite{Bozza:2002zj}. For null geodesics in static and spherically symmetric spacetimes, this structure is well established~\cite{Bozza:2002zj,Tsukamoto:2016jzh} (see also, e.g., Refs.~\cite{Perlick:2004,Bozza:2010xqn} for reviews). The coefficients $\bar{a}$ and $\bar{b}$ encode the leading strong-field behavior and determine the main lensing observables. This logarithmic divergence implies an infinite sequence of higher-order relativistic images~\cite{Ohanian:1987xga,Virbhadra:1999nm} and admits finite-distance extensions~\cite{Ishihara:2016sfv,Ono:2019hkw}.

An analogous SDL arises for the scattering of massive particles following timelike geodesics. For a fixed specific energy $E\ge 1$, a near-critical particle approaches an $E$-dependent unstable circular orbit, and the deflection angle diverges logarithmically in the Schwarzschild spacetime~\cite{Tsupko:2014wza,Liu:2015zou} and the Reissner--Nordstr\"om spacetime~\cite{Pang:2018jpm}. Related near-critical or logarithmic behavior also appears in ultrarelativistic particle lensing (e.g., neutrinos)~\cite{Mena:2006ym,Eiroa:2008ks,Taak:2022xdp} and near-separatrix zoom-whirl dynamics in extreme-mass-ratio inspirals~\cite{Amaro-Seoane:2012lgq,Glampedakis:2002ya,Barack:2019agd}. Similar strong-deflection behavior also appears in dispersive photon propagation in a plasma~\cite{Tsupko:2013cqa,Perlick:2017fio,Bisnovatyi-Kogan:2017kii,Matsuno:2020kju}. Recently, Ref.~\cite{Feleppa:2024kio} generalized this behavior to generic static and spherically symmetric spacetimes.

Recent work has shown that the SDL coefficients for photons admit local, coordinate-invariant expressions, in particular, the logarithmic coefficient $\bar{a}$ can be written in terms of curvature data on the unstable circular photon orbit~\cite{Igata:2025taz,Igata:2025plb}. This unstable circular photon orbit is closely connected to eikonal quasinormal modes (QNMs) in many black hole spacetimes~\cite{Ferrari:1984zz,Cardoso:2008bp}. However, many instability-based interpretations are formulated in terms of effective potentials and employ a Lyapunov exponent defined with respect to the Killing time~\cite{Stefanov:2010xz,Raffaelli:2014ola}---this exponent is related to the SDL coefficient $\bar{a}$.

Despite recent progress toward local, coordinate-invariant characterizations of the SDL coefficients for photons, the geometric origin and physical interpretation of the SDL coefficients for massive particles remain less transparent. Existing derivations of the SDL for massive particles---including the general framework of Ref.~\cite{Feleppa:2024kio}---typically extract the logarithmic divergence via coordinate-based near-critical expansions of the deflection-angle integral. What is still missing is a derivation that (i) identifies the leading logarithmic coefficient with a locally defined instability rate obtained directly from the geodesic deviation equation and (ii) provides a curvature-based, local characterization of the energy-dependent critical timelike trajectory, in close analogy with the null case~\cite{Igata:2025taz,Igata:2025plb}. Related analyses of gravitational lensing based on the geodesic deviation equation have recently been carried out for null congruences in the Schwarzschild spacetime under the weak-deflection, thin-lens approximation within geometric optics~\cite{Li:2024oke}. In contrast, the present work concerns timelike near-critical scattering in generic static, spherically symmetric spacetimes and shows that the leading logarithmic coefficient in the strong-deflection expansion is controlled by the radial instability exponent of the critical trajectory.

To fill this gap, we treat the critical timelike trajectory as the fundamental local ingredient governing strong-deflection scattering. We formulate the SDL for massive particles in terms of kinematic and geometric data intrinsic to that trajectory---most notably the locally measured orbital speed $v_{\mathrm{c}}$, the areal radius $R_{\mathrm{c}}$, and tidal curvature components in a suitable orthonormal frame. These orbit-local data define a radial instability exponent $\kappa_{\mathrm{c}}$ via the growth rate of radial deviations governed by the geodesic deviation equation, and $\kappa_{\mathrm{c}}$ in turn controls the logarithmic divergence of the deflection angle in the strong-deflection regime.

The present curvature-based formulation is also related to geometrical approaches to gravitational lensing in which optical scalars and deflection angles are expressed directly in terms of curvature scalars and matter fields. In the weak-deflection regime, the optical scalars and the deflection angle can be written in terms of projected Ricci and Weyl curvature scalars, and, for spherically symmetric lenses, in terms of the energy-momentum tensor components~\cite{Gallo:2011mv}. This curvature-based formulation was extended to second order in weak lensing~\cite{Crisnejo:2017jmx}. It was also generalized to dispersive media, including cold plasma, and a relation to the deflection of massive charged particles was discussed using optical-metric and Jacobi-metric methods~\cite{Crisnejo:2019xtp}. While these works share with ours the aim of characterizing lensing through geometrical or matter-field data, they mainly concern weak-deflection lensing and, in the dispersive-media case, frequency-dependent propagation or nongeodesic motion. The present work instead develops a strong-deflection formulation for timelike geodesics and gives a local dynamical interpretation of the logarithmic SDL coefficient in terms of the radial instability exponent of the critical orbit.

The present analysis is complementary to Ref.~\cite{Igata:2026ivq}, which develops a strong-deflection expansion for null geodesics near the marginally unstable circular photon orbit associated with a degenerate photon sphere, where the deflection angle exhibits a power-law rather than logarithmic divergence in the SDL (see also Refs.~\cite{Tsukamoto:2025hbz,Sasaki:2025web}). By contrast, the present paper treats massive particles following timelike geodesics near a nondegenerate unstable circular orbit. The key new step here is to show covariantly, via the geodesic deviation equation, that the logarithmic SDL coefficient is governed by the radial instability exponent of that critical timelike trajectory, and to express this exponent in terms of local kinematic and curvature data.

Our main results are as follows: (i) a unified SDL expansion for massive-particle scattering with a smooth high-energy limit, (ii) a covariant derivation, from the geodesic deviation equation, of the key relation $\bar{a}=1/\kappa_{\mathrm{c}}$ for $E>1$ [Eq.~\eqref{eq:result}] and, equivalently, $a=2/\kappa_{\mathrm{c}}$ for $E\ge1$ [Eq.~\eqref{eq:result2}], (iii) curvature-based expressions for $\kappa_{\mathrm{c}}$ in the comoving and static frames, and (iv) in general relativity, a compact matter dependence of $\kappa_{\mathrm{c}}$ (and hence $\bar{a}$) through a single local scalar built from the static-frame energy density and principal pressures. We also emphasize that $\bar{a}$ is determined entirely by local data on the unstable circular orbit, whereas $\bar{b}$ contains a nonlocal contribution through the regular part of the deflection integral.

This paper is organized as follows. In Sec.~\ref{sec:2}, we formulate the deflection angle for timelike geodesics in general static and spherically symmetric spacetimes. In Sec.~\ref{sec:3}, we derive the conditions for energy-dependent unstable circular orbits and characterize their instability, identifying the critical trajectory. In Sec.~\ref{sec:4}, we define the SDL for massive particles and obtain the logarithmic divergence of the deflection angle. In Sec.~\ref{sec:5}, we clarify the local curvature origin of the logarithmic divergence coefficient by relating it to curvature components and to the instability exponent of the critical trajectory. In Sec.~\ref{sec:6}, we discuss the high-energy limit of the logarithmic coefficient. In Sec.~\ref{sec:7}, we analyze the effect of matter content on the SDL for massive particles. In Sec.~\ref{sec:8}, we relate the energy-dependent SDL coefficients to strong-lensing observables, including relativistic image positions, angular separations, flux ratios, and time delays. We also discuss how these observables can be used to reconstruct the critical impact parameter and the instability exponent. Finally, in Sec.~\ref{sec:9}, we summarize our results, discuss their implications and limitations, and outline possible extensions. Throughout this paper, we use geometrized units with $G=1$ and $c=1$ and employ abstract index notation~\cite{Wald:1984} when convenient.

%%%%%
\section{Deflection angle of massive particles in static and spherically symmetric spacetimes}
\label{sec:2}
%%%%%
We consider a general static and spherically symmetric spacetime described by the line element
\begin{align}
\mathrm{d}s^2=-A(r)\:\!\mathrm{d}t^2+B(r)\:\!\mathrm{d}r^2+R(r)^2 (\mathrm{d}\theta^2+\sin^2\theta\:\!\mathrm{d}\varphi^2).
\end{align}
The function $R(r)$ is the areal radius, so that a 2-sphere at constant $t$ and $r$ has area $4\pi R(r)^2$. We work in the static region where $A(r)>0$ and $B(r)>0$, assume asymptotic flatness ($A,B\to1$ and $R\to r$ as $r\to \infty$), and restrict to the domain connected to infinity with $R'(r)>0$ so that $r$ is in one-to-one correspondence with $R$. Hereafter, a prime denotes differentiation with respect to $r$.

To analyze gravitational deflection, we model the trajectories of massive particles as timelike geodesics. By spherical symmetry, the motion can be restricted, without loss of generality, to the equatorial plane $\theta=\pi/2$. The geodesic equations follow from the Lagrangian
\begin{align}
\mathscr{L}
=\frac{1}{2}\bigl[\:\!-A(r)\:\!\dot{t}^{\:\!2}+B(r)\:\!\dot{r}^{\:\!2}+R(r)^2\dot{\varphi}^{\:\!2}\:\!\bigr],
\end{align}
where the overdot denotes differentiation with respect to the proper time along the worldline. In a static, spherically symmetric spacetime, there are two conserved quantities: the specific energy $E= A(r)\:\!\dot{t}$ (i.e., the energy per unit rest mass) and the specific angular momentum $L=R(r)^2 \dot{\varphi}$.

The normalization condition for timelike geodesics, $2\mathscr{L}=-1$, combined with the conserved quantities $E$ and $L$, gives the radial equation of motion
\begin{align}
\dot{r}^{\:\!2}+\frac{L^2-H(r)}{B(r)R(r)^2}=0, 
\qquad H(r)\equiv R(r)^2\left(\frac{E^2}{A(r)}-1\right).
\label{eq:radeq}
\end{align}
For nonradial motion ($L\neq0$), using the relation $\dot r=(\mathrm{d}r/\mathrm{d}\varphi)\dot\varphi$ together with $L = R^2 \dot{\varphi}$, we can rewrite Eq.~\eqref{eq:radeq} as the orbital equation
\begin{align}
\left(
\frac{\mathrm{d}r}{\mathrm{d}\varphi}
\right)^2+V(r)=0,
\label{eq:orbitaleq}
\end{align}
where the effective potential $V(r)$ is defined by
\begin{align}
V(r)=\frac{R(r)^2}{B(r)}\left(1-\frac{H(r)}{L^2}\right).
\label{eq:V}
\end{align}
Since $(\mathrm{d}r/\mathrm{d}\varphi)^2 \ge 0$, the motion is restricted to the allowed region where $V(r)\le0$.

We focus on scattering trajectories with $E\ge 1$ (unbound for $E>1$ and marginally bound for $E=1$), in which a particle arrives from infinity, reaches a turning point (point of closest approach), and returns to infinity. The radial turning point is determined by $V_0\equiv V(r_0)=0$. Hereafter, the subscript $0$ denotes evaluation at $r=r_0$. Using the explicit form of $V(r)$ in Eq.~\eqref{eq:V}, this condition can be rewritten as
\begin{align}
L^2=R_0^2\left(
\frac{E^2}{A_0}-1
\right).
\end{align}
Using this relation, we eliminate $L$ and parametrize the trajectory in terms of the turning point $R_0$ (equivalently, $r_0$). The effective potential then takes the form
\begin{align}
V(r)=\frac{R(r)^2}{B(r)}\left(
1-\frac{R(r)^2}{R_0^2}\frac{A_0}{A(r)}\frac{1-\frac{A(r)}{E^2}}{1-\frac{A_0}{E^2}}
\right).
\end{align}

It is convenient to introduce the local speed measured by static observers. A static observer with four-velocity $\bar{e}_{(0)}=A^{-1/2}\partial_t$ measures the Lorentz factor $\gamma(r)=E/\sqrt{A(r)}$ (so $\gamma\to E$ at infinity), and the corresponding three-speed $v(r)=\sqrt{1-\gamma(r)^{-2}}$ satisfies
\begin{align}
v(r)^2 =1-\frac{A(r)}{E^2}.
\label{eq:v2}
\end{align}
This expression shows that $v(r)\to 1$ in the high-energy limit $E\to\infty$ (at fixed $r$ in the static region), so timelike geodesics approach null geodesics, and $V(r)$ reduces to the null case. We return to this limit in Sec.~\ref{sec:6}. In terms of $v(r)$, the effective potential takes the form
\begin{align}
V(r)=\frac{R(r)^2}{B(r)}\left(
1-\frac{R(r)^2}{R_0^2}\frac{A_0}{A(r)}\frac{v(r)^2}{v_0^2}
\right).
\label{eq:Vv}
\end{align}

From Eq.~\eqref{eq:orbitaleq}, the total change of the azimuthal angle along a scattering trajectory is given by
\begin{align}
\Delta \varphi(R_0, E)=2\int_{r_0(R_0)}^\infty \frac{\mathrm{d}r}{\sqrt{-V(r)}},
\label{eq:totalchange}
\end{align}
where $V(r)<0$ for $r> r_0$. The deflection angle for a source and an observer at infinity is 
\begin{align}
\hat{\alpha}(R_0, E)=\Delta\varphi(R_0, E)-\pi.
\label{eq:deflectionangle}
\end{align}

%%%%%
\section{Unstable circular orbits of massive particles}
\label{sec:3}
%%%%%
We derive the conditions for the $E$-dependent unstable circular timelike orbit and its radial instability exponent, thereby identifying the critical trajectory that controls the SDL.

The conditions for circular orbits follow directly from Eq.~\eqref{eq:orbitaleq}. A circular orbit at $r=r_{\mathrm{c}}$ satisfies $V(r_{\mathrm{c}})=0$ and $V'(r_{\mathrm{c}})=0$ simultaneously. Using the explicit form of $V(r)$ in Eq.~\eqref{eq:V}, the condition~$V_{\mathrm{c}}=0$ yields 
\begin{align}
L^2=L_{\mathrm{c}}^2\equiv R_{\mathrm{c}}^2\left(
\frac{E^2}{A_{\mathrm{c}}}-1
\right),
\label{eq:circ1}
\end{align}
where $A_{\mathrm{c}}\equiv A(r_{\mathrm{c}})$ and $R_{\mathrm{c}}\equiv R(r_{\mathrm{c}})$. Hereafter, the subscript $\mathrm{c}$ denotes evaluation at $r=r_{\mathrm{c}}$. 
Using Eq.~\eqref{eq:circ1} to eliminate $L$, the condition $V'_{\mathrm{c}}=0$ gives
\begin{align}
\frac{A'_{\mathrm{c}}}{A_{\mathrm{c}}}-\frac{2R'_{\mathrm{c}}}{R_{\mathrm{c}}}\left(1-\frac{A_{\mathrm{c}}}{E^2}\right)=0. 
\label{eq:ccond}
\end{align}
Equation~\eqref{eq:ccond} determines the $E$-dependent circular-orbit radius $r_{\mathrm{c}}(E)$ [equivalently, $R_{\mathrm{c}}(E)$], and Eq.~\eqref{eq:circ1} then gives $L_{\mathrm{c}}(E)$. For $L_{\mathrm{c}}\neq 0$, Eq.~\eqref{eq:circ1} implies $0<v_{\mathrm{c}}^2=1-A_{\mathrm{c}}/E^2<1$, and Eq.~\eqref{eq:ccond} can be rewritten as
\begin{align}
v_{\mathrm{c}}^2=1-\frac{A_{\mathrm{c}}}{E^2}=\frac{1}{2}\frac{R_{\mathrm{c}}A'_{\mathrm{c}}}{R'_{\mathrm{c}}A_{\mathrm{c}}}.
\label{eq:vccond}
\end{align}

To characterize the radial stability, we consider a small radial deviation from the circular orbit $R(\varphi)=R_{\mathrm{c}} \bigl[1+\tilde{\delta}(\varphi)\bigr]$, with $|\tilde{\delta}|\ll1$, at fixed $E$ and $L=L_{\mathrm{c}}(E)$. Using $\mathrm{d}r/\mathrm{d}\varphi=(1/R')(\mathrm{d}R/\mathrm{d}\varphi)$, Eq.~\eqref{eq:orbitaleq} can be written in terms of $R$ as 
\begin{align}
\left(
\frac{\mathrm{d}R}{\mathrm{d}\varphi}
\right)^2+(R')^2\:\!V(r)=0.
\label{eq:orbitaleqR}
\end{align}
Expanding $(R')^2V$ about $\tilde{\delta}=0$ and using $V_{\mathrm{c}}=0$ and $V'_{\mathrm{c}}=0$, we find $(R')^2V= (R_{\mathrm{c}}^2/2)V''_{\mathrm{c}} \tilde{\delta}^2+O(\tilde{\delta}^3)$. Substituting this expansion into Eq.~\eqref{eq:orbitaleqR} yields, to leading order,
\begin{align}
\biggl(
\frac{\mathrm{d}\tilde{\delta}}{\mathrm{d}\varphi}
\biggr)^2-\kappa_{\mathrm{c}}^2 \:\!\tilde{\delta}^2\simeq 0,
\end{align}
where $\kappa_{\mathrm{c}}^2\equiv -V''_{\mathrm{c}}/2$. The sign of $V''_{\mathrm{c}}$ (and hence that of $\kappa_{\mathrm{c}}^2$) determines radial stability. For the unstable branch relevant to the SDL, $V''_{\mathrm{c}}<0$ so that $\kappa_{\mathrm{c}}^2>0$, and we take 
\begin{align}
\kappa_{\mathrm{c}}=\sqrt{-\frac{V''_{\mathrm{c}}}{2}}>0.
\label{eq:kappaV''c}
\end{align}
The perturbation then admits exponentially growing or decaying solutions
\begin{align}
\tilde{\delta}\propto \exp(\pm \kappa_{\mathrm{c}} \varphi),
\label{eq:zeta}
\end{align}
confirming the radial instability.%
\footnote{On the stable branch ($V''_{\mathrm{c}}>0$), radial perturbations lead to epicyclic oscillations (see Sec.~\ref{sec:5B}). The marginal case $V''_{\mathrm{c}}=0$ can yield nonlogarithmic (power-law) behavior~\cite{Igata:2026ivq}.} Hereafter, we focus on the critical trajectory at $L=L_{\mathrm{c}}$, obtained as the limit of scattering trajectories with $L\to L_{\mathrm{c}}^{+}$. This trajectory asymptotically approaches the unstable circular orbit rather than returning to infinity.

We now derive explicit expressions for $\kappa_{\mathrm{c}}^2$ (equivalently, $V''_{\mathrm{c}}$). Eliminating $L$ via Eq.~\eqref{eq:circ1} and using Eq.~\eqref{eq:ccond} to eliminate $A'_{\mathrm{c}}$, we find 
\begin{align}
\kappa_{\mathrm{c}}^2
=\frac{R_{\mathrm{c}}^2}{B_{\mathrm{c}}}\left[\:\!
\frac{R''_{\mathrm{c}}}{R_{\mathrm{c}}}+\left(1-\frac{4A_{\mathrm{c}}}{E^2}\right)\left(
\frac{R'_{\mathrm{c}}}{R_{\mathrm{c}}}
\right)^2-\left(1-\frac{A_{\mathrm{c}}}{E^2}\right)^{-1}
 \frac{A''_{\mathrm{c}}}{2A_{\mathrm{c}}}
\:\!\right].
\label{eq:V''m1}
\end{align}
Using Eq.~\eqref{eq:vccond}, this can be reexpressed in terms of the locally measured speed $v_{\mathrm{c}}$ as
\begin{align}
\kappa_{\mathrm{c}}^2
=\frac{R_{\mathrm{c}}^2}{B_{\mathrm{c}}}\left[\:\!
\frac{R''_{\mathrm{c}}}{R_{\mathrm{c}}}+(4v_{\mathrm{c}}^2-3)\left(
\frac{R'_{\mathrm{c}}}{R_{\mathrm{c}}}
\right)^2-\frac{1}{2v_{\mathrm{c}}^2} \frac{A''_{\mathrm{c}}}{A_{\mathrm{c}}}
\:\!\right].
\label{eq:V''m2}
\end{align}

We summarize the local condition for the outer turning point relevant to scattering from infinity. The allowed region $V(r)\le 0$ is equivalent to $L^2\le H(r)$, and a turning point $r_0$ satisfies $L^2=H_0$. The critical value $L_{\mathrm{c}}$ is reached when the turning-point radius coincides with the radius $r_{\mathrm{c}}$ of the unstable circular orbit, i.e.,
\begin{align}
H_{\mathrm{c}}=L_{\mathrm{c}}^2,
\qquad
H'_{\mathrm{c}}=0, \qquad 
H''_{\mathrm{c}}>0,
\end{align}
so $r_{\mathrm{c}}$ is a local minimum of $H(r)$. Accordingly, the branch with an outer turning point $r_0>r_{\mathrm{c}}$ exists for $L^2>L_{\mathrm{c}}^2$. The case $L=L_{\mathrm{c}}$ marks the separatrix between trajectories that turn back outside $r_{\mathrm{c}}$ and trajectories that penetrate to $r<r_{\mathrm{c}}$. For $L^2<L_{\mathrm{c}}^2$, the subsequent motion is determined by the global structure of $H(r)$ and may include plunge or additional turning behavior, depending on the spacetime.

%%%%%
\section{Deflection angle in the strong deflection limit}
\label{sec:4}
%%%%%
In this section, we define the SDL for massive particles as the critical limit in which the turning-point radius of a scattering trajectory approaches the radius of the $E$-dependent unstable circular orbit. The deflection angle then diverges logarithmically, and its leading coefficient is fixed by the local behavior of $V(r)$ at the unstable circular orbit, in particular, by $V''_{\mathrm{c}}$.

We consider scattering trajectories with specific energy $E\ge 1$. Scattering trajectories have a turning point $r_0(E,L)$ determined by $V_0=0$, while the unstable circular orbit at $r=r_{\mathrm{c}}(E)$ defines the critical angular momentum $L_{\mathrm{c}}(E)$. The SDL is defined by approaching the critical angular momentum from the scattering side $L\to L_{\mathrm{c}}(E)^{+}$, equivalently $r_0\to r_{\mathrm{c}}(E)^{+}$ or $R_0\to R_{\mathrm{c}}(E)^{+}$.

We analyze the behavior of the deflection angle~\eqref{eq:deflectionangle} in the SDL. We rewrite the integral of Eq.~\eqref{eq:totalchange} using the variable $z$ introduced in Ref.~\cite{Igata:2025taz}, 
\begin{align}
z=1-\frac{R_0}{R},
\end{align}
which measures the deviation from the turning point in terms of the areal radius $R$. The turning point corresponds to $z=0$, while $R\to\infty$ is mapped to $z\to1$.%
\footnote{This definition of $z$ (based on the areal radius $R$) differs from the variables commonly used, e.g., in Refs.~\cite{Bozza:2002zj,Tsukamoto:2016jzh}.}
We hold $E$ fixed and suppress the explicit dependence on $E$ when convenient. Using $\mathrm{d}r=\mathrm{d}R/R'$ and $\mathrm{d}R=R_0\mathrm{d}z/(1-z)^2$, the total azimuthal advance~\eqref{eq:totalchange} can be rewritten as
\begin{align}
\Delta \varphi(R_0)
=2 \int_0^1 f(z; R_0)\:\! \mathrm{d}z,
\label{eq:IR0}
\end{align}
where, for later convenience, the integrand $f(z; R_0)$ is defined by
\begin{align}
f(z; R_0)=\left[\:\!(1-z)^4\frac{(R')^2 (-V)}{R_0^2}\:\!\right]^{-1/2}.
\label{eq:f}
\end{align}

In the SDL, the divergence is controlled by the integrand near the turning point $z=0$. Expanding the argument of the square root in $f(z;R_0)$ about $z=0$ [using $V(r_0)=0$] yields
\if0
\begin{align}
R'&=R'_0+\frac{R_0 R''_0}{R'_0}z+O(z^2),
\\
V&=\frac{R_0 V'_0}{R'_0} z+\left[\:\!\left(
\frac{R_0}{R'_0}-\frac{R_0^2 R''_0}{2(R'_0)^3}
\right)V'_0
+\frac{R_0^2}{2(R'_0)^2} V''_0\:\!\right]z^2+O(z^3),
\end{align}
\fi
\begin{align}
f(z; R_0)=\left[\:\!c_1 z+c_2 z^2+O(z^3)\:\!\right]^{-1/2},
\end{align}
where
\begin{align}
c_1&=-\frac{R'_0 V'_0}{R_0},
\\
c_2&=3\left(
\frac{R'_0}{R_0}-\frac{R''_0}{2R'_0}
\right)V'_0-\frac{V''_0}{2}.
\end{align}
For $c_1\neq0$ (i.e., $V'_0\neq0$), the integrand behaves as $f(z; R_0)\sim z^{-1/2}$ and the integral remains finite. In the SDL, $R_0\to R_{\mathrm{c}}$ implies $V'_0\to0$ and hence $c_1\to 0$, so the leading behavior becomes $f(z; R_0)\sim z^{-1}$, and $\Delta \varphi(R_0)$ diverges logarithmically. We isolate the divergent part by defining
\begin{align}
I_{\mathrm{D}}(R_0)=2\int_0^1 \frac{\mathrm{d}z}{\sqrt{c_1 z+c_2 z^2}}.
\label{eq:ID}
\end{align}
For $c_2>0$ and $c_1> 0$, Eq.~\eqref{eq:ID} yields
\begin{align}
I_{\mathrm{D}}(R_0)
&=\frac{4}{\sqrt{c_2}}
\log\!\left(
\frac{\sqrt{c_2}+\sqrt{c_1+c_2}}{\sqrt{c_1}}
\right).
\label{eq:ID_closed}
\end{align}

We next expand the coefficients $c_1$ and $c_2$ around $R_0=R_{\mathrm{c}}(E)$. For a fixed energy $E$, we introduce the dimensionless deviation defined by
\begin{align}
\delta = \frac{R_0}{R_{\mathrm{c}}}-1,
\label{eq:delta}
\end{align}
where $\delta>0$ for scattering trajectories. In the SDL, as $\delta\to0^+$, the coefficients admit the asymptotic expansions
\begin{align}
c_1=-V''_{\mathrm{c}}\delta+O(\delta^2),
\qquad
c_2= -\frac{V''_{\mathrm{c}}}{2} +O(\delta),
\label{eq:csexp}
\end{align}
where $V''_{\mathrm{c}}<0$ for an unstable circular orbit. Substituting Eq.~\eqref{eq:csexp} into Eq.~\eqref{eq:ID_closed} and expanding about $\delta =0$, we obtain 
\begin{align}
I_{\mathrm{D}}(R_0)=-a \log \delta+a \log 2+O(\delta \log \delta),
\end{align}
where $a$ is determined by $V''_{\mathrm{c}}$ as
\begin{align}
a&=2\sqrt{-\frac{2}{V''_{\mathrm{c}}}}.
\label{eq:SDLa}
\end{align}

We write $\Delta \varphi(R_0)=I_{\mathrm{D}}(R_0)+I_{\mathrm{R}}(R_0)$ with
\begin{align}
I_{\mathrm{R}}(R_0)=2\int_0^1 
\mathrm{d}z\:\!\left[\:\!
f(z; R_0)-\frac{1}{\sqrt{c_1 z+c_2 z^2}}
\:\!\right].
\label{eq:IR}
\end{align}
By construction, the integrand in Eq.~\eqref{eq:IR} is finite at $z=0$ in the SDL, so $I_{\mathrm{R}}(R_0)$ approaches a finite limit. We define the regular contribution to the constant term as 
\begin{align}
b_{\mathrm{R}}\equiv \lim_{R_0\to R_{\mathrm{c}}} I_{\mathrm{R}}(R_0),
\label{eq:SDLbR}
\end{align}
which encodes a nonlocal (global) contribution from the spacetime geometry.

Combining the divergent and regular parts, the deflection angle in the SDL admits an expansion of the universal form%
\footnote{Reference~\cite{Feleppa:2024kio} quotes an $O(\delta)$ remainder for $I_{\mathrm{D}}$ [see Eq.~(48) therein]. Retaining the $O(\delta)$ correction to the logarithmic prefactor yields an $O(\delta \log \delta)$ remainder, corresponding to $O(\varepsilon^{1/2} \log \varepsilon)$ in Eq.~\eqref{eq:alphaimp}.}
\begin{align}
\hat{\alpha}(R_0, E)=-a \log \delta+b+O(\delta \log \delta),
\label{eq:alphadelta}
\end{align}
where $a$ is given in Eq.~\eqref{eq:SDLa}, and the constant term $b$ is defined by
\begin{align}
b=a \log 2+b_{\mathrm{R}}-\pi. 
\label{eq:betadelta}
\end{align}
Here, $a$ is determined by local data on the circular orbit (through $V''_{\mathrm{c}}$), whereas $b$ contains the nonlocal term $b_{\mathrm{R}}$ [Eq.~\eqref{eq:SDLbR}].

We reexpress Eq.~\eqref{eq:alphadelta} in terms of the asymptotic impact parameter. For $E>1$, we define%
\footnote{
In the marginally bound case $E=1$, $\eta=L/\sqrt{E^2-1}$ is ill defined, so we use the turning-point representation $R_0\to R_{\mathrm{c}}$.}
\begin{align}
\eta\equiv \frac{L}{\sqrt{E^2-1}}.
\end{align}
In the high-energy limit, $\eta\to L/E$, so $\eta$ coincides with the impact parameter for a photon. The critical value is obtained by evaluating $\eta$ at $R_0=R_{\mathrm{c}}$, 
\begin{align}
\eta_{\mathrm{c}}=\frac{R_{\mathrm{c}}}{\sqrt{A_{\mathrm{c}}}}\,
\sqrt{\frac{E^2-A_{\mathrm{c}}}{E^2-1}}.
\label{eq:etac}
\end{align}
Expanding $\eta$ about $\delta=0$ yields
\begin{align}
\eta=\eta_{\mathrm{c}}\left[\:\!
1-\frac{B_{\mathrm{c}} V''_{\mathrm{c}}}{4(R'_{\mathrm{c}})^2} \delta^2+O(\delta^3)
\:\!\right].
\label{eq:etaexp}
\end{align}
The absence of an $O(\delta)$ term reflects that $\eta$ is extremized at $R_0=R_{\mathrm{c}}$.

To quantify the fractional deviation from criticality, we introduce 
\begin{align}
\varepsilon=\frac{\eta}{\eta_{\mathrm{c}}}-1.
\end{align}
For scattering trajectories sufficiently close to the critical trajectory, where $0<\varepsilon\ll1$, Eq.~\eqref{eq:etaexp} can be perturbatively inverted to give
\begin{align}
\delta=\sqrt{-\frac{4(R'_{\mathrm{c}})^2}{V''_{\mathrm{c}}B_{\mathrm{c}}}}\:\!\varepsilon^{1/2}+O(\varepsilon). 
\label{eq:delep}
\end{align}
We rewrite $(R'_{\mathrm{c}})^2/B_{\mathrm{c}}$ in terms of the Misner--Sharp mass~\cite{Misner:1964je,Hayward:1994bu,Kinoshita:2024wyr}, defined as 
\begin{align}
m(r)
\equiv \frac{R(r)}{2}\Bigl[1-g^{ab}(\nabla_a R)(\nabla_b R)\Bigr]
=\frac{R(r)}{2}\left(1-\frac{(R')^2}{B(r)}\right).
\label{eq:MSmass}
\end{align}
On the unstable circular orbit, this gives $(R'_{\mathrm{c}})^2/B_{\mathrm{c}} = 1-2m_{\mathrm{c}}/R_{\mathrm{c}}$, where $m_{\mathrm{c}}=m(r_{\mathrm{c}})$. Combining this with Eq.~\eqref{eq:delep}, we obtain 
\begin{align}
\hat{\alpha}(\eta, E)=-\bar{a} \log \varepsilon+\bar{b}+O(\varepsilon^{1/2}\log \varepsilon),
\label{eq:alphaimp}
\end{align}
where 
\begin{align}
\bar{a}&=\frac{a}{2}=\sqrt{-\frac{2}{V''_{\mathrm{c}}}},
\label{eq:abar}
\\
\bar{b}&=-\pi+b_{\mathrm{R}}-\bar{a} \log \left[\:\!
\frac{\bar{a}^2}{2}\left(1-\frac{2m_{\mathrm{c}}}{R_{\mathrm{c}}}\right)
\:\!\right].
\label{eq:bbar}
\end{align}
The expansion has the same form as in the photon case, but the coefficients now depend on $E$: $\bar{a}$ through $V''_{\mathrm{c}}$, and $\bar{b}$ through $V''_{\mathrm{c}}$, $m_{\mathrm{c}}$, and $b_{\mathrm{R}}$. In the common domain $E>1$, Eqs.~\eqref{eq:alphadelta} and~\eqref{eq:alphaimp} do not represent different physical deflection laws. They are the same SDL expansion of the deflection angle for a single near-critical scattering family, expressed in terms of different criticality parameters: the turning-point deviation $\delta$ and the impact-parameter deviation $\varepsilon$, which are related by Eq.~\eqref{eq:delep}. The turning-point representation is nevertheless useful because it is geometrically transparent and remains well defined also for the marginally bound case $E=1$, where $\eta$ is ill defined.

For $E>1$, the SDL coefficients obtained above agree with the results in Ref.~\cite{Feleppa:2024kio} upon identifying our asymptotic impact parameter $\eta$ with their $u$. Note that the intermediate steps are organized differently: we define $\delta$ geometrically via the areal radius [Eq.~\eqref{eq:delta}], whereas the analysis of Ref.~\cite{Feleppa:2024kio} employs a different, more coordinate-based near-critical parametrization and expansion scheme. Therefore, intermediate quantities such as $b_{\mathrm{R}}$ are scheme dependent and need not agree between the two formulations.

The result in Eq.~\eqref{eq:abar} can be written, for $E>1$, in terms of the quantity $\kappa_{\mathrm{c}}$ introduced in Eq.~\eqref{eq:kappaV''c}, as
\begin{align}
\bar{a}=\frac{1}{\kappa_{\mathrm{c}}}.
\label{eq:abarkappac}
\end{align}
In the turning-point representation,%
\footnote{
Near the unstable circular orbit, the radial deviation $\tilde{\delta}$ decays exponentially with the azimuthal angle as in Eq.~\eqref{eq:zeta}, $\tilde{\delta} \propto e^{-\kappa_{\mathrm{c}}\varphi}$. Hence, $\varphi\sim -\kappa_{\mathrm{c}}^{-1} \log |\tilde{\delta}|$, and since $\Delta \varphi\gg \pi$ in the SDL, the divergent part of the deflection behaves as $\hat{\alpha}\simeq \Delta \varphi \sim -(2/\kappa_{\mathrm{c}})\log |\tilde{\delta}|$.}
we likewise obtain, for $E\geq 1$, 
\begin{align}
a=\frac{2}{\kappa_{\mathrm{c}}}. 
\label{eq:akappac}
\end{align}
In Sec.~\ref{sec:5B}, we show that $\kappa_{\mathrm{c}}$ coincides with the radial characteristic exponent obtained from the geodesic deviation equation, thus providing a covariant interpretation of Eqs.~\eqref{eq:abarkappac} and \eqref{eq:akappac}.

For practical computation, first solve for the critical trajectory $r_{\mathrm{c}}(E)$ [Eqs.~\eqref{eq:ccond} or~\eqref{eq:vccond}]. For $E>1$, determine $\eta_{\mathrm{c}}$ [Eq.~\eqref{eq:etac}], evaluate $\kappa_{\mathrm{c}}$ [Eq.~\eqref{eq:V''m1} or~\eqref{eq:V''m2}] to obtain $\bar{a}$, and use $b_{\mathrm{R}}$ [Eq.~\eqref{eq:SDLbR}] and $m_{\mathrm{c}}$ [Eq.~\eqref{eq:MSmass}] in Eq.~\eqref{eq:bbar} to obtain $\bar{b}$. In the marginally bound case $E=1$, the near-critical expansion should instead be expressed in terms of the turning-point deviation $\delta$ [Eq.~\eqref{eq:alphadelta}], with coefficients $a$ [Eq.~\eqref{eq:SDLa}] and $b$ [Eq.~\eqref{eq:betadelta}].

%%%%%
\section{Local curvature origin of strong deflection}
\label{sec:5}
%%%%%
Motivated by the relation between the logarithmic divergence coefficients and $\kappa_{\mathrm{c}}$, we now show that $\kappa_{\mathrm{c}}$ admits a covariant interpretation in terms of local curvature data evaluated on the unstable circular orbit. This section proceeds in two steps. In Sec.~\ref{sec:5A}, we express the instability exponent $\kappa_{\mathrm{c}}$ in terms of curvature components in an orthonormal frame comoving with the circular orbit and then rewrite the result in the static frame. In Sec.~\ref{sec:5B}, we use the geodesic deviation equation to provide a covariant characterization of the instability of the circular orbit and to identify the relevant characteristic exponent with $\kappa_{\mathrm{c}}$, thereby justifying its role as the exponent governing the logarithmic divergence of the deflection angle.

%%%
\subsection{Curvature formulation in a comoving orthonormal frame}
\label{sec:5A}
%%%
We introduce a comoving orthonormal tetrad $\{e_{(\mu)}\}$ on the circular orbit, with $e_{(0)}$ aligned with the particle's four-velocity,
\begin{align}
e_{(0)}&=\frac{1}{\alpha_{\mathrm{c}}} \left(\partial_t+\Omega_{\mathrm{c}} \partial_\varphi \right), 
\label{eq:e0}
\\
e_{(1)}&=\frac{1}{\sqrt{B_{\mathrm{c}}}} \partial_r,
\\
e_{(2)}&=\frac{1}{R_{\mathrm{c}}} \partial_\theta,
\\
e_{(3)}&=
\frac{1}{\alpha_{\mathrm{c}}}\left(
v_{\mathrm{c}} \partial_t+
\frac{\Omega_{\mathrm{c}}}{v_{\mathrm{c}}}
\partial_\varphi
\right),
\label{eq:e3}
\end{align}
where $\alpha_{\mathrm{c}}=\sqrt{A_{\mathrm{c}}}/\gamma_{\mathrm{c}}$ denotes the redshift factor on the circular orbit. The coordinate angular velocity on the orbit is
\begin{align}
\Omega_{\mathrm{c}}=\frac{v_{\mathrm{c}}\sqrt{A_{\mathrm{c}}}}{R_{\mathrm{c}}},
\label{eq:Omegac}
\end{align}
where we used Eqs.~\eqref{eq:circ1} and~\eqref{eq:vccond}.

The electric part of the Weyl tensor $C_{abcd}$ measured by the comoving observer is $E_{(i)(j)} = C_{abcd}e^a_{(i)}e^b_{(0)}e^c_{(j)}e^d_{(0)}$. Similarly, the tetrad components of the Einstein tensor $G_{ab}$ are defined by $G_{(\mu)(\nu)} = G_{ab}\, e_{(\mu)}^{a} e_{(\nu)}^{b}$. The label $\mathrm{c}$ on curvature quantities such as $G_{(\mu)(\nu)}^{\:\!\mathrm{c}}$ and $E_{(i)(j)}^{\:\!\mathrm{c}}$ indicates evaluation on the unstable circular orbit $r=r_{\mathrm{c}}$, with all tensor components projected onto the comoving tetrad.

Evaluating the comoving frame curvature components on the unstable circular orbit and using Eq.~\eqref{eq:vccond}, we obtain
\begin{align}
\kappa_{\mathrm{c}}^2
=\frac{R_{\mathrm{c}}^2}{\gamma_{\mathrm{c}}^2v_{\mathrm{c}}^2}\left[\:\!
-E_{(1)(1)}^{\:\! \mathrm{c}}-3E_{(3)(3)}^{\:\! \mathrm{c}}
-\frac{1}{6}\left(
4\:\!G_{(0)(0)}^{\:\! \mathrm{c}}
+5\:\!G_{(1)(1)}^{\:\! \mathrm{c}}
+8\:\!G_{(2)(2)}^{\:\! \mathrm{c}}
-G_{(3)(3)}^{\:\! \mathrm{c}}
\right)
\:\!\right].
\label{eq:kappaEG}
\end{align}
This expresses $\kappa_{\mathrm{c}}^2$ (and hence $\bar{a}^{-2}$) in terms of a linear combination of curvature components evaluated on the circular orbit, entering through the dimensionless combinations $R_{\mathrm{c}}^2 E_{(i)(i)}^{\mathrm{c}}$ and $R_{\mathrm{c}}^2 G_{(\mu)(\mu)}^{\mathrm{c}}$, together with the kinematic prefactor $(\gamma_{\mathrm{c}}v_{\mathrm{c}})^{-2}$.

In the high-energy limit $v_{\mathrm{c}}\to1$ ($\gamma_{\mathrm{c}}\to\infty$), the comoving tetrad degenerates because $e_{(0)}$ and $e_{(3)}$ become parallel---this is a frame artifact. We, therefore, introduce a static frame and the boost relations for later use.

%%%
\subsubsection{Static frame and boost relations}
\label{sec:5A1}
%%%
We introduce an orthonormal tetrad associated with static observers at $r=r_{\mathrm{c}}$ on the equatorial plane
\begin{align}
\bar{e}_{(0)}=\frac{1}{\sqrt{A_{\mathrm{c}}}}\partial_t,
\qquad
\bar{e}_{(1)}=\frac{1}{\sqrt{B_{\mathrm{c}}}}\partial_r,
\qquad
\bar{e}_{(2)}=\frac{1}{R_{\mathrm{c}}}\partial_\theta,
\qquad
\bar{e}_{(3)}=\frac{1}{R_{\mathrm{c}}}\partial_\varphi.
\label{eq:staticframe}
\end{align}
We denote tensor components measured in this frame by an overbar. On the circular orbit, the comoving tetrad~\eqref{eq:e0}--\eqref{eq:e3} is related to the static frame by a Lorentz boost in the azimuthal direction with relative speed $v_{\mathrm{c}}$,
\begin{align}
e_{(0)}
=\gamma_{\mathrm{c}}\bigl(\bar e_{(0)}+v_{\mathrm{c}}\bar e_{(3)}\bigr),
\qquad
e_{(1)}=\bar{e}_{(1)},
\qquad
e_{(2)}=\bar{e}_{(2)},
\qquad
e_{(3)}
=\gamma_{\mathrm{c}}\bigl(v_{\mathrm{c}}\bar{e}_{(0)}+\bar{e}_{(3)}\bigr).
\label{eq:boost}
\end{align}

By staticity and spherical symmetry, $\bar{G}^{\:\!\mathrm{c}}_{(\mu)(\nu)}$ is diagonal and the Weyl tensor is purely electric and diagonal in the static frame, with $\bar{G}^{\:\!\mathrm{c}}_{(2)(2)}=\bar{G}^{\:\!\mathrm{c}}_{(3)(3)}$ and $\bar{E}^{\:\!\mathrm{c}}_{(2)(2)}=\bar{E}^{\:\!\mathrm{c}}_{(3)(3)}$. The trace-free condition $\bar{E}^{\:\!\mathrm{c}}_{(1)(1)}+2\bar{E}^{\:\!\mathrm{c}}_{(2)(2)}=0$ then gives
\begin{align}
\bar{E}^{\:\!\mathrm{c}}_{(i)(j)}=\mathrm{diag}\left(
-2\bar{\mathcal{E}}_{\mathrm{c}},\, \bar{\mathcal{E}}_{\mathrm{c}},\, \bar{\mathcal{E}}_{\mathrm{c}}
\right),
\end{align}
where $\bar{\mathcal{E}}_{\mathrm{c}}\equiv \bar E^{\mathrm{c}}_{(2)(2)}$.

With these properties, the boost~\eqref{eq:boost} yields simple transformation laws for the curvature components entering Eq.~\eqref{eq:kappaEG}. For the electric part of the Weyl tensor, the nonzero diagonal components transform as
\begin{align}
E^{\:\!\mathrm{c}}_{(1)(1)} 
&
=-(2+v_{\mathrm{c}}^2)\gamma_{\mathrm{c}}^2 \,\bar{\mathcal{E}}_{\mathrm{c}},
\label{eq:E11boost}
\\
E^{\:\!\mathrm{c}}_{(2)(2)} 
&
=(1+2v_{\mathrm{c}}^2)\gamma_{\mathrm{c}}^2\, \bar{\mathcal{E}}_{\mathrm{c}},
\\
E^{\:\!\mathrm{c}}_{(3)(3)} &= \bar{E}^{\:\!\mathrm{c}}_{(3)(3)}=\bar{\mathcal{E}}_{\mathrm{c}}.
\label{eq:E33boost}
\end{align}
Similarly, the diagonal components of the Einstein tensor transform as follows:
\begin{align}
G^{\:\!\mathrm{c}}_{(0)(0)} &= \gamma_{\mathrm{c}}^2\!\left(\bar G^{\:\!\mathrm{c}}_{(0)(0)} + v_{\mathrm{c}}^2\,\bar G^{\mathrm{c}}_{(3)(3)}\right), 
\label{eq:G00boost}
\\
G^{\:\!\mathrm{c}}_{(3)(3)} &= \gamma_{\mathrm{c}}^2\!\left(\bar{G}^{\:\!\mathrm{c}}_{(3)(3)}+v_{\mathrm{c}}^2\,\bar{G}^{\:\!\mathrm{c}}_{(0)(0)} \right),
\\
G^{\:\!\mathrm{c}}_{(1)(1)} &= \bar{G}^{\:\!\mathrm{c}}_{(1)(1)}, 
\quad
G^{\:\!\mathrm{c}}_{(2)(2)} = \bar{G}^{\:\!\mathrm{c}}_{(2)(2)}.
\label{eq:G11G22boost}
\end{align}
Substituting Eqs.~\eqref{eq:E11boost}--\eqref{eq:E33boost} and Eqs.~\eqref{eq:G00boost}--\eqref{eq:G11G22boost} into Eq.~\eqref{eq:kappaEG}, we obtain
\begin{align}
\kappa_{\mathrm{c}}^2
=\frac{R_{\mathrm{c}}^2}{v_{\mathrm{c}}^2} \left[\:\!
\left(4v_{\mathrm{c}}^2-1\right) \bar{\mathcal{E}}_{\mathrm{c}}
-\frac{4-v_{\mathrm{c}}^2}{6}\left(
\bar{G}^{\:\!\mathrm{c}}_{(0)(0)}
+\bar{G}^{\:\!\mathrm{c}}_{(3)(3)}\right)
-\frac{1-v_{\mathrm{c}}^2}{6}\left(5\bar{G}^{\:\!\mathrm{c}}_{(1)(1)}+3\bar{G}^{\:\!\mathrm{c}}_{(3)(3)}\right)
\:\!\right].
\label{eq:kappacstatic}
\end{align}

Using the circular-orbit condition~\eqref{eq:vccond}, we eliminate $\bar{\mathcal{E}}_{\mathrm{c}}$ in favor of static-frame Einstein tensor components via
\begin{align}
R_{\mathrm{c}}^2\bar{\mathcal{E}}_{\mathrm{c}}=
\frac{v_{\mathrm{c}}^2}{1+2v_{\mathrm{c}}^2}-\frac{R_{\mathrm{c}}^2}{6}\left[
\bar{G}^{\:\!\mathrm{c}}_{(0)(0)}+\bar{G}^{\:\!\mathrm{c}}_{(3)(3)}
+\frac{2(1-v_{\mathrm{c}}^2)}{1+2v_{\mathrm{c}}^2}\bar{G}^{\:\!\mathrm{c}}_{(1)(1)}
\right]. 
\label{eq:Ebar_in_terms_of_Gbar}
\end{align}
Substituting Eq.~\eqref{eq:Ebar_in_terms_of_Gbar} into Eq.~\eqref{eq:kappacstatic}, we find
\begin{align}
\kappa_{\mathrm{c}}^2
=
\frac{4v_{\mathrm{c}}^2-1}{2v_{\mathrm{c}}^2+1}
-\frac{R_{\mathrm{c}}^2}{v_{\mathrm{c}}^2}\left[\:\!
\frac{1+v_{\mathrm{c}}^2}{2}\bar{G}^{\:\!\mathrm{c}}_{(0)(0)}
+\bar{G}^{\:\!\mathrm{c}}_{(3)(3)}
+\frac{(1-v_{\mathrm{c}}^2)(6v_{\mathrm{c}}^2+1)}{2(2v_{\mathrm{c}}^2+1)}\bar{G}^{\:\!\mathrm{c}}_{(1)(1)}
\:\!\right],
\label{eq:kappa_Gbar_only}
\end{align}
which depends only on $R_{\mathrm{c}}$, $v_{\mathrm{c}}$, and the static-frame Einstein tensor components. We use Eq.~\eqref{eq:kappa_Gbar_only} in Secs.~\ref{sec:6} and~\ref{sec:7}.

We rewrite the factor $(1-2m_{\mathrm{c}}/R_{\mathrm{c}})$ in Eq.~\eqref{eq:bbar} using local static-frame quantities,
\begin{align}
1-\frac{2m_{\mathrm{c}}}{R_{\mathrm{c}}}
=\frac{1+R_{\mathrm{c}}^2\bar{G}^{\:\!\mathrm{c}}_{(1)(1)}}{1+2v_{\mathrm{c}}^2}. 
\end{align}
Substituting this relation into Eq.~\eqref{eq:bbar}, we obtain an equivalent representation of the SDL constant term,
\begin{align}
\bar{b}=-\pi+b_{\mathrm{R}}-\bar{a} \log \left(
\frac{\bar{a}^2}{2}\frac{1+R_{\mathrm{c}}^2\bar{G}^{\:\!\mathrm{c}}_{(1)(1)}}{1+2v_{\mathrm{c}}^2}
\right). 
\end{align}
Apart from the nonlocal term $b_{\mathrm{R}}$, $\bar{b}$ is fixed by local data on the circular orbit.

%%%%%
\subsection{Geodesic deviation: A covariant derivation of the instability exponent}
\label{sec:5B}
%%%%%
Using the geodesic deviation equation, we identify the radial characteristic exponent of the circular timelike geodesic with $\kappa_{\mathrm{c}}$, thereby providing a covariant origin of $\bar{a}=1/\kappa_{\mathrm{c}}$.

We consider a timelike geodesic congruence with tangent $u^a$ ($u^a u_a=-1$) and an orthogonal deviation vector $\xi^a$ ($u_a\xi^a=0$) that is Lie-transported along $u^a$. The deviation obeys the geodesic deviation equation%
\footnote{Related curvature-based stability analyses for null geodesics and photon surfaces include Refs.~\cite{Koga:2019uqd,Claudel:2000yi,Yoshino:2016kgi,Koga:2020akc}.}
\begin{align}
u^c \nabla_c (u^b \nabla_b \:\!\xi^a)+R_{cbd}{}^a u^c u^d \xi^b=0,
\end{align}
where $R_{cbd}{}^a$ is the Riemann curvature tensor. Decomposing it into the Weyl tensor $C_{cbd}{}^a$ and the trace parts built from the Ricci tensor $R_{ab}$ and the scalar curvature $\mathcal{R}$ yields 
\begin{align}
u^c \nabla_c(u^b \nabla_b \:\!\xi^a)+C_{cbd}{}^a u^c u^d \xi^b-\frac{1}{2}R^a{}_b\xi^b-\frac{1}{2} (R_{bc}u^b\xi^c)\:\! u^a+\frac{1}{2}(R_{bc}u^b u^c)\:\!\xi^a+\frac{\mathcal{R}}{6} \xi^a=0.
\label{eq:devC}
\end{align}

We introduce a comoving orthonormal tetrad $e_{(\mu)}^a$ with $e_{(0)}^a=u^a$, and denote the spatial components of the deviation vector by $\xi^{(i)} = \delta^{(i)(j)}e_{(j)a} \xi^{a}$, where $\delta^{(i)(j)}$ is the Kronecker delta.

We define the Ricci rotation coefficients by
\begin{align}
\omega_{(i)(j)}=e_{(i)a}u^b \nabla_b e_{(j)}^a,
\end{align}
which satisfy $\omega_{(i)(j)}=-\omega_{(j)(i)}$. In the comoving tetrad~\eqref{eq:e0}--\eqref{eq:e3}, Eq.~\eqref{eq:devC} becomes
\begin{align}
\ddot{\xi}^{(i)}+2\omega^{(i)}{}_{(j)} \dot{\xi}^{(j)}+\left(
\dot{\omega}^{(i)}{}_{(j)}+\omega^{(i)}{}_{(k)} \omega^{(k)}{}_{(j)}+\mathcal{T}^{(i)}{}_{(j)}
\right) \xi^{(j)}=0,
\label{eq:deveqtet}
\end{align}
where a dot denotes differentiation with respect to $\tau$, and spatial tetrad indices $(i), (j), \ldots$ are raised and lowered with the Kronecker delta. The curvature terms $\mathcal{T}^{(i)}{}_{(j)}$ represent the tidal effects experienced by the congruence and are defined as
\begin{align}
\mathcal{T}^{(i)}{}_{(j)}=E^{(i)}{}_{(j)}-\frac{1}{2}G^{(i)}{}_{(j)}+\left(
\frac{1}{2}G_{(0)(0)}-\frac{\mathcal{R}}{3}
\right)\delta^{(i)}{}_{(j)},
\label{eq:tidaltensor}
\end{align}
where the curvature components are those introduced in Sec.~\ref{sec:5A}. The scalar curvature obeys the trace relation $\mathcal{R}=G_{(0)(0)}-G_{(1)(1)}-G_{(2)(2)}-G_{(3)(3)}$.

For the equatorial circular orbit [using the tetrad \eqref{eq:e0}--\eqref{eq:e3}], all coefficients are constant and $\varphi$ is linear in $\tau$. Using $\mathrm{d}/\mathrm{d}\tau=\dot{\varphi}_{\mathrm{c}}\,\mathrm{d}/\mathrm{d}\varphi$, Eq.~\eqref{eq:deveqtet} becomes
\begin{align}
\frac{\mathrm{d}^2\xi^{(i)}}{\mathrm{d}\varphi^2}
+2\tilde{\omega}^{(i)}{}_{(j)} \frac{\mathrm{d}\xi^{(j)}}{\mathrm{d}\varphi}
+\left(
\tilde{\omega}^{(i)}{}_{(k)} \tilde{\omega}^{(k)}{}_{(j)}+\tilde{\mathcal{T}}^{(i)}{}_{(j)}
\right) \xi^{(j)}=0,
\label{eq:devphi}
\end{align}
where 
\begin{align}
\tilde{\omega}^{(i)}{}_{(j)}&\equiv 
\frac{1}{\dot{\varphi}_{\mathrm{c}}}\bigl(\:\!\omega^{(i)}{}_{(j)}\:\!\bigr)_{\mathrm{c}},
\\
\tilde{\mathcal{T}}^{(i)}{}_{(j)} &\equiv \frac{1}{\dot{\varphi}_{\mathrm{c}}^2}
\bigl(\:\!\mathcal{T}^{(i)}{}_{(j)}\:\!\bigr)_{\mathrm{c}},
\end{align}
and the only nonvanishing components of $\tilde{\omega}^{(i)}{}_{(j)}$ are given by
\begin{align}
\tilde{\omega}^{(1)}{}_{(3)}
=-\tilde{\omega}^{(3)}{}_{(1)}
=-\frac{1}{\dot{\varphi}_{\mathrm{c}}}\frac{R'_{\mathrm{c}}\Omega_{\mathrm{c}}}{\sqrt{A_{\mathrm{c}}B_{\mathrm{c}}}}\equiv \tilde{\omega}.
\end{align}

Because Eq.~\eqref{eq:devphi} has constant coefficients, it admits mode solutions $\xi^{(i)}(\varphi)\propto \exp (\kappa\varphi)$ with a characteristic exponent $\kappa$. Substituting this ansatz into Eq.~\eqref{eq:devphi} gives the eigenvalue problem
\begin{align}
\left(
\kappa^2 \delta^{(i)}{}_{(j)}+2\kappa\:\! \tilde{\omega}^{(i)}{}_{(j)}+\tilde{\mathcal{T}}_{\mathrm{eff}}^{(i)}{}_{(j)}
\right) \xi^{(j)}=0,
\label{eq:eigeneq}
\end{align}
where the effective tidal matrix is
\begin{align}
\tilde{\mathcal{T}}_{\mathrm{eff}}^{(i)}{}_{(j)}\equiv \tilde{\mathcal{T}}^{(i)}{}_{(j)}+\tilde{\omega}^{(i)}{}_{(k)} \tilde{\omega}^{(k)}{}_{(j)}
=\mathrm{diag}\left(\tilde{\mathcal{T}}^{(1)}{}_{(1)}-\tilde{\omega}^2,\tilde{\mathcal{T}}^{(2)}{}_{(2)},\tilde{\mathcal{T}}^{(3)}{}_{(3)}-\tilde{\omega}^2\right).
\end{align}

The axial Killing vector $\partial_\varphi$ generates a trivial deviation mode with $\kappa=0$ aligned with $e_{(3)}^a$. Substituting $\kappa=0$ into Eq.~\eqref{eq:eigeneq} gives 
\begin{align}
\tilde{\mathcal{T}}^{(3)}{}_{(3)}=\tilde{\omega}^2,
\label{eq:zeromodeid}
\end{align}
which encodes the existence of the trivial azimuthal shift (gauge) mode.%
\footnote{A displacement generated by $\partial_\varphi$ can be made orthogonal to $u^a$ using the gauge freedom $\xi^a\to \xi^a+f(\varphi)\, u^a$ with $f = u_a\partial_\varphi^a$, which is conserved. The resulting deviation is $\varphi$ independent and aligned with $e_{(3)}^a$.} 
Nontrivial modes satisfy the characteristic equation
\begin{align}
\det\left(
\kappa^2 \delta^{(i)}{}_{(j)}+2\kappa\:\! \tilde{\omega}^{(i)}{}_{(j)}+\tilde{\mathcal{T}}_{\mathrm{eff}}^{(i)}{}_{(j)}
\right)=0.
\label{eq:det}
\end{align}
Combining Eqs.~\eqref{eq:det} and~\eqref{eq:zeromodeid} yields
\begin{align}
&\kappa_{(1)}^2=-\tilde{\mathcal{T}}^{(1)}{}_{(1)}-3\tilde{\mathcal{T}}^{(3)}{}_{(3)},
\label{eq:prekappa1}
\\
&\kappa_{(2)}^2=-\tilde{\mathcal{T}}^{(2)}{}_{(2)}.
\label{eq:prekappa2}
\end{align}
The exponent $\kappa_{(1)}$ governs radial deviations (and hence the SDL): $\kappa_{(1)}^2>0$ corresponds to radial instability, whereas $\kappa_{(1)}^2<0$ gives epicyclic oscillations in $\varphi$ with frequency $\sqrt{-\kappa_{(1)}^2}$.%
\footnote{The second exponent $\kappa_{(2)}$ governs vertical (out-of-plane) stability and is not used below.}

Substituting Eq.~\eqref{eq:tidaltensor} into Eqs.~\eqref{eq:prekappa1} and~\eqref{eq:prekappa2}, we express these exponents explicitly in terms of curvature components evaluated on the circular orbit,
\begin{align}
\kappa_{(1)}^2&=
\frac{R_{\mathrm{c}}^2}{\gamma_{\mathrm{c}}^2v_{\mathrm{c}}^2}
\left[\:\!
-E_{(1)(1)}^{\:\! \mathrm{c}}-3E_{(3)(3)}^{\:\! \mathrm{c}}
-\frac{1}{6}\left(
4\:\!G_{(0)(0)}^{\:\! \mathrm{c}}
+5\:\!G_{(1)(1)}^{\:\! \mathrm{c}}
+8\:\!G_{(2)(2)}^{\:\! \mathrm{c}}
-G_{(3)(3)}^{\:\! \mathrm{c}}
\right)
\:\!\right],
\label{eq:kappa1}
\\
\kappa_{(2)}^2&
=
\frac{R_{\mathrm{c}}^2}{\gamma_{\mathrm{c}}^2v_{\mathrm{c}}^2}
\left[\:\!-E_{(2)(2)}^{\:\! \mathrm{c}}
-\frac{1}{6}\left(
G_{(0)(0)}^{\:\! \mathrm{c}}
+2G_{(1)(1)}^{\:\! \mathrm{c}}
-G_{(2)(2)}^{\:\! \mathrm{c}}
+2G_{(3)(3)}^{\:\! \mathrm{c}}
\right)
\:\!\right],
\end{align}
where we have used Eqs.~\eqref{eq:circ1} and~\eqref{eq:vccond} to write $\dot{\varphi}_{\mathrm{c}}$ as
\begin{align}
\dot{\varphi}_{\mathrm{c}}^2
=\frac{\gamma_{\mathrm{c}}^2v_{\mathrm{c}}^2}{R_{\mathrm{c}}^2}.
\end{align}
Each exponent separates into a kinematic prefactor and a local curvature combination.

Comparing Eq.~\eqref{eq:kappa1} with Eq.~\eqref{eq:kappaEG}, we find
\begin{align}
\kappa_{(1)}^2=\kappa_{\mathrm c}^2.
\end{align}
On the unstable branch relevant to the SDL, we take the positive root and obtain
\begin{align}
\kappa_{(1)}=\kappa_{\mathrm c}>0.
\label{eq:kappaid}
\end{align}
This provides a covariant interpretation of the quantity $\kappa_{\mathrm c}$ introduced through the local expansion of the effective potential: it is precisely the radial characteristic exponent of the critical orbit with respect to the azimuthal angle.

Combining Eq.~\eqref{eq:kappaid} with Eq.~\eqref{eq:abarkappac}, we then obtain, for $E>1$, 
\begin{align}
\bar{a}=\frac{1}{\kappa_{\mathrm{c}}}=\frac{1}{\kappa_{(1)}}.
\label{eq:result}
\end{align}
In the turning-point representation, combining Eq.~\eqref{eq:kappaid} with Eq.~\eqref{eq:akappac}, we obtain, for $E\ge1$, 
\begin{align}
a=\frac{2}{\kappa_{\mathrm c}}=\frac{2}{\kappa_{(1)}}.
\label{eq:result2}
\end{align}

%%%%%
\section{High-energy limit}
\label{sec:6}
%%%%%
In this section, we take the high-energy limit $E\to\infty$, in which the critical timelike trajectory approaches the corresponding null critical trajectory, while the associated unstable circular timelike orbit approaches the circular photon orbit. Since the comoving frame introduced in Sec.~\ref{sec:5A} degenerates in this limit, we use the static-frame expression for $\kappa_{\mathrm{c}}$ (and hence $\bar{a}=1/\kappa_{\mathrm{c}}$).

Taking $E\to\infty$ in Eq.~\eqref{eq:ccond} yields the standard circular photon orbit condition (see, e.g., Ref.~\cite{Bozza:2002zj}),
\begin{align}
\frac{A'_{\mathrm{p}}}{A_{\mathrm{p}}}-\frac{2R'_{\mathrm{p}}}{R_{\mathrm{p}}}=0, 
\label{eq:ccondUR}
\end{align}
where the label $\mathrm{p}$ denotes evaluation on the circular photon orbit $r=r_{\mathrm{p}}=\lim_{E\to \infty}r_{\mathrm{c}}(E)$.
Combining Eq.~\eqref{eq:ccondUR} with Eq.~\eqref{eq:vccond} gives $v_{\mathrm{c}}^2\to1$ in this limit, consistent with the null character of the orbit.

Next, we take the $E\to \infty$ limit of Eq.~\eqref{eq:V''m1} and obtain
\begin{align}
\kappa_{\mathrm{p}}^2
=\frac{R_{\mathrm{p}}^2}{B_{\mathrm{p}}}\left[\:\!
\frac{R''_{\mathrm{p}}}{R_{\mathrm{p}}}+\left(
\frac{R'_{\mathrm{p}}}{R_{\mathrm{p}}}
\right)^2-
 \frac{A''_{\mathrm{p}}}{2A_{\mathrm{p}}}
\:\!\right].
\end{align}
Taking $v_{\mathrm{c}}\to 1$ in Eq.~\eqref{eq:kappa_Gbar_only} yields
\begin{align}
\kappa_{\mathrm{p}}^2
= 
1-R_{\mathrm{p}}^2\bigl(\bar{G}^{\:\!\mathrm{p}}_{(0)(0)}+\bar{G}^{\:\!\mathrm{p}}_{(3)(3)}\bigr).
\label{eq:kappap}
\end{align}
The corresponding SDL coefficient in the high-energy limit is then 
\begin{align}
\bar{a}=\frac{1}{\sqrt{1-R_{\mathrm{p}}^2\bigl(\bar{G}^{\:\!\mathrm{p}}_{(0)(0)}+\bar{G}^{\:\!\mathrm{p}}_{(3)(3)}\bigr)}}.
\label{eq:ap}
\end{align}
This expression reproduces the SDL coefficient for photons derived in Ref.~\cite{Igata:2025taz} (see also Ref.~\cite{Igata:2025plb} for a Newman--Penrose description).

%%%%%
\section{Matter contributions}
\label{sec:7}
%%%%%
Specializing to general relativity, we use the Einstein equations to express the matter dependence of $\kappa_{\mathrm{c}}$ (and hence $\bar{a}$). In the static frame~\eqref{eq:staticframe}, the Einstein equations read $\bar{G}_{(\mu)(\nu)} = 8\pi \bar{T}_{(\mu)(\nu)}$. The stress-energy tensor is diagonal, $\bar{T}_{(\mu)(\nu)}=\mathrm{diag}(\rho,\,P,\,\Pi,\,\Pi)$, where $\rho$ is the energy density, $P$ is the radial pressure, and $\Pi$ is the tangential pressure measured by static observers.

Using the Einstein equations, Eq.~\eqref{eq:kappa_Gbar_only} can be rewritten in terms of the matter variables as 
\begin{align}
\kappa_{\mathrm{c}}^2
=
\frac{4v_{\mathrm{c}}^2-1}{2v_{\mathrm{c}}^2+1}
-\frac{4\pi R_{\mathrm{c}}^2}{v_{\mathrm{c}}^2(2v_{\mathrm{c}}^2+1)}\mathcal{S}_{\mathrm{c}},
\label{eq:kappac_matter}
\end{align}
where
\begin{align}
\mathcal{S}_{\mathrm{c}}=(2v_{\mathrm{c}}^2+1)\left[\:\!
(1+v_{\mathrm{c}}^2)\rho_{\mathrm{c}}
+2\Pi_{\mathrm{c}}
\:\!\right]
+(1-v_{\mathrm{c}}^2)(6v_{\mathrm{c}}^2+1)
P_{\mathrm{c}}.
\label{eq:Sc1}
\end{align}
Equation~\eqref{eq:kappac_matter} shows that the matter contribution is proportional to $-\mathcal{S}_{\mathrm{c}}$: $\mathcal{S}_{\mathrm{c}}>0$ decreases $\kappa_{\mathrm{c}}^2$ (increasing $\bar{a}$), while $\mathcal{S}_{\mathrm{c}}<0$ increases $\kappa_{\mathrm{c}}^2$ (decreasing $\bar{a}$). For $0<v_{\mathrm{c}}^2<1$, all coefficients in Eq.~\eqref{eq:Sc1} are non-negative, so $\rho_{\mathrm{c}},P_{\mathrm{c}},\Pi_{\mathrm{c}}\ge0$ implies $\mathcal{S}_{\mathrm{c}}\ge0$.

To connect $\mathcal{S}_{\mathrm{c}}$ to standard energy condition combinations, we rewrite it as 
\begin{align}
\mathcal{S}_{\mathrm{c}}=
(1-v_{\mathrm{c}}^2)(1+4v_{\mathrm{c}}^2)(\rho_{\mathrm{c}}+P_{\mathrm{c}}+2\Pi_{\mathrm{c}})
+2v_{\mathrm{c}}^2 (1-v_{\mathrm{c}}^2)(\rho_{\mathrm{c}}+P_{\mathrm{c}})
+2v_{\mathrm{c}}^2(4v_{\mathrm{c}}^2-1)(\rho_{\mathrm{c}}+\Pi_{\mathrm{c}}).
\label{eq:Sc2}
\end{align}
In Eq.~\eqref{eq:Sc2}, the first two coefficients are non-negative for $0<v_{\mathrm{c}}^2<1$, while the coefficient of $(\rho_{\mathrm{c}}+\Pi_{\mathrm{c}})$ flips sign at $v_{\mathrm{c}}^2=1/4$ (i.e., $v_{\mathrm{c}}=1/2$). If the strong energy condition (SEC) holds on the circular orbit ($\rho_{\mathrm{c}}+P_{\mathrm{c}}\ge0$, $\rho_{\mathrm{c}}+\Pi_{\mathrm{c}}\ge0$, and $\rho_{\mathrm{c}}+P_{\mathrm{c}}+2\Pi_{\mathrm{c}}\ge0$), then $\mathcal{S}_{\mathrm{c}}\ge0$ is guaranteed for $v_{\mathrm{c}}^2\ge 1/4$. For $v_{\mathrm{c}}^2<1/4$, the SEC alone is inconclusive without additional assumptions on anisotropy.

For an isotropic perfect fluid $P_{\mathrm{c}}=\Pi_{\mathrm{c}}\equiv p_{\mathrm{c}}$, Eq.~\eqref{eq:Sc2} reduces to 
\begin{align}
\mathcal{S}_{\mathrm{c}}=
(1-v_{\mathrm{c}}^2)(1+4v_{\mathrm{c}}^2)(\rho_{\mathrm{c}}+3p_{\mathrm{c}})
+6v_{\mathrm{c}}^4(\rho_{\mathrm{c}}+p_{\mathrm{c}}),
\end{align}
so the perfect-fluid SEC (i.e., $\rho_{\mathrm{c}}+p_{\mathrm{c}}\ge0$ and $\rho_{\mathrm{c}}+3p_{\mathrm{c}}\ge0$) implies $\mathcal{S}_{\mathrm{c}}\ge0$. By contrast, a vacuum energy-like equation of state $p_{\mathrm{c}}=-\rho_{\mathrm{c}}$ yields $\mathcal{S}_{\mathrm{c}}<0$ for $\rho_{\mathrm{c}}>0$, which increases $\kappa_{\mathrm{c}}^2$ and decreases $\bar{a}$.

For anisotropic stresses with $\rho_{\mathrm{c}}>0$, we define $w_{\mathrm{r}}\equiv P_{\mathrm{c}}/\rho_{\mathrm{c}}$ and $w_{\mathrm{t}}\equiv \Pi_{\mathrm{c}}/\rho_{\mathrm{c}}$. Equation~\eqref{eq:Sc1} then becomes
\begin{align}
\frac{\mathcal{S}_{\mathrm{c}}}{\rho_{\mathrm{c}}}
=(2v_{\mathrm{c}}^2+1)\left(
1+v_{\mathrm{c}}^2+2w_{\mathrm{t}}\right)
+(1-v_{\mathrm{c}}^2)(6v_{\mathrm{c}}^2+1)\,w_{\mathrm{r}}.
\end{align}
For a fixed $v_{\mathrm{c}}$, the condition $\mathcal{S}_{\mathrm{c}}=0$ is the straight line
\begin{align}
w_{\mathrm{r}}=-\frac{(2v_{\mathrm{c}}^2+1)(
1+v_{\mathrm{c}}^2+2w_{\mathrm{t}})}
{(1-v_{\mathrm{c}}^2)(6v_{\mathrm{c}}^2+1)}
\end{align}
in the $(w_{\mathrm{t}},w_{\mathrm{r}})$ plane (with $\mathcal{S}_{\mathrm{c}}>0$ above it). The null energy condition enforces $w_{\mathrm{r}}, w_{\mathrm{t}}\ge -1$ but leaves the sign of $\mathcal{S}_{\mathrm{c}}$ unconstrained. Examples include a radial Maxwell field $(w_{\mathrm{r}},w_{\mathrm{t}})=(-1,1)$ (giving $\mathcal{S}_{\mathrm{c}}>0$), the vacuum energy point $(-1,-1)$ (giving $\mathcal{S}_{\mathrm{c}}<0$), and radial tension $(-1,0)$, for which $\mathcal{S}_{\mathrm{c}}=2v_{\mathrm{c}}^2(4v_{\mathrm{c}}^2-1)\rho_{\mathrm{c}}$ exhibits the sign flip at $v_{\mathrm{c}}^2=1/4$.

In vacuum ($T_{ab}=0$), $\mathcal{S}_{\mathrm{c}}=0$ and Eq.~\eqref{eq:kappac_matter} reduces to
\begin{align}
\kappa_{\mathrm{c}}^2=\frac{4v_{\mathrm{c}}^2-1}{2v_{\mathrm{c}}^2+1},
\end{align}
implying that radial instability requires $v_{\mathrm{c}}^2>1/4$. For an asymptotically flat vacuum spacetime, Birkhoff's theorem implies the Schwarzschild metric, $A(r)=1-2M/r$ and $B(r)=1/A(r)$ [with areal radius $R(r)=r$]. Using Eq.~\eqref{eq:vccond}, we obtain
$v_{\mathrm{c}}^2=M/[r_{\mathrm{c}}(E)-2M]$, and hence
$\kappa_{\mathrm{c}}^2=[6M-r_{\mathrm{c}}(E)]/r_{\mathrm{c}}(E)$ and
\begin{align}
\bar{a}=\sqrt{\frac{r_{\mathrm{c}}(E)}{6M-r_{\mathrm{c}}(E)}},
\end{align}
with $r_{\mathrm{c}}(E)$ from 
\begin{align}
\frac{r_{\mathrm{c}}(E)}{M}=\frac{3E^2-4+E \sqrt{9E^2-8}}{2(E^2-1)}.
\end{align}
This agrees with the results of Ref.~\cite{Feleppa:2024kio}.%
\footnote{Introducing a parameter $x\in [1/3,1)$ by $E=\sqrt{8/[9(1-x^2)]}$, we obtain $\bar{a}=\sqrt{(1+x)/(2x)}$, in agreement with Ref.~\cite{Feleppa:2024kio}. }
More generally, the same local relations apply whenever an open neighborhood of the critical trajectory is vacuum and spherically symmetric, since the metric there is locally isometric to Schwarzschild (up to a constant rescaling of the Killing time).

Finally, we consider the high-energy limit of Eq.~\eqref{eq:kappac_matter}. As $E\to \infty$, the critical timelike orbit approaches the circular photon orbit, $r_{\mathrm{c}} \to r_{\mathrm{p}}$ and $v_{\mathrm{c}}^2\to 1$. Thus, we obtain
\begin{align}
\kappa_{\mathrm{p}}^2=1-8\pi R_{\mathrm{p}}^2(\rho_{\mathrm{p}}+\Pi_{\mathrm{p}})
\end{align}
and the corresponding leading SDL coefficient
\begin{align}
\bar{a}=\frac{1}{\sqrt{1-8\pi R_{\mathrm{p}}^2(\rho_{\mathrm{p}}+\Pi_{\mathrm{p}})}},
\end{align}
where the subscript $\mathrm{p}$ denotes evaluation on the circular photon orbit. These expressions are the matter-field representations of Eqs.~\eqref{eq:kappap} and~\eqref{eq:ap} in general relativity, respectively. In this limit, the SDL coefficient is controlled only by the tangential null energy combination $\rho_{\mathrm{p}}+\Pi_{\mathrm{p}}$, provided that the circular photon orbit is nondegenerate and unstable.

%%%%%
\section{Strong-lensing observables and detectability}
\label{sec:8}
%%%%%
We now express the critical impact parameter and the SDL coefficients in terms of standard strong-lensing observables, including the limiting angular position, the angular separation of the relativistic images, their flux ratio, and the winding contribution to the time delay at leading order. This construction follows the standard strong-deflection lensing formalism developed for
photons in Ref.~\cite{Bozza:2002zj}. Here, we extend it to fixed-energy massive particles by
replacing the photon critical impact parameter and SDL coefficients with the energy-dependent
quantities $\eta_{\mathrm{c}}(E)$, $\bar{a}(E)$, and $\bar{b}(E)$. 

Let $D_{\mathrm{OL}}$ denote the distance between the observer $\mathrm{O}$ and the lens center $\mathrm{L}$, and let $\vartheta$ denote the angular separation of an observed image with respect to the optical axis $\mathrm{OL}$. Throughout this section, we fix the particle energy $E>1$ and consider a single relativistic image branch on one side of the optical axis, so that $\vartheta>0$---the branch on the opposite side is obtained by reflection. In the asymptotic observer region, and within the small-angle approximation, the impact parameter $\eta$ of the detected trajectory is related to $\vartheta$ by $\eta=D_{\mathrm{OL}} \sin \vartheta \simeq D_{\mathrm{OL}}\vartheta$. We, therefore, use the impact-parameter form of the SDL expansion, Eq.~\eqref{eq:alphaimp}, for the observational quantities considered below, while Eq.~\eqref{eq:alphadelta} gives an equivalent turning-point representation. Introducing the critical angular position $\vartheta_{\mathrm{c}}(E)
\simeq
\eta_{\mathrm{c}}(E)/D_{\mathrm{OL}}$, we rewrite the SDL expansion~\eqref{eq:alphaimp} as
\begin{align}
\hat{\alpha}(\vartheta,E)
\simeq
-\bar a(E)
\log \varepsilon+\bar b(E), \qquad 
\varepsilon=\frac{\vartheta}{\vartheta_{\mathrm{c}}(E)}-1.
\label{eq:angular-SDL}
\end{align}

For a trajectory in the $n$-th winding branch, we separate the integer multiple of $2\pi$ from the deflection angle by defining
\begin{align}
\Delta \alpha_n(\vartheta, E)=\hat{\alpha}-2n\pi, \qquad n=1, 2, \ldots
\end{align}
Near the $n$-th relativistic image, $\Delta\alpha_n$ is treated as a small residual deflection. We introduce the reference angular position $\vartheta=\vartheta_n^0$ of the $n$-th relativistic image branch by
\begin{align}
\Delta\alpha_n(\vartheta_n^0,E)=0.
\label{eq:vartheta-n0-def}
\end{align}
We parametrize its small offset from the critical angular position by
\begin{align}
\varepsilon_n(E)=\frac{\vartheta_n^0(E)}{\vartheta_{\mathrm c}(E)}-1,
\label{eq:vepn}
\end{align}
so that 
\begin{align}
\vartheta_n^0-\vartheta_{\mathrm c}=\vartheta_{\mathrm c} \varepsilon_n.
\label{eq:thetacepn}
\end{align}
Using the SDL form~\eqref{eq:angular-SDL} in Eq.~\eqref{eq:vartheta-n0-def}, we obtain
\begin{align}
\varepsilon_n
=
\exp\left(
\frac{\bar b(E)-2\pi n}{\bar a(E)}
\right).
\label{eq:epsilon-n}
\end{align}
Thus, the zeroth-order relativistic image positions approach
$\vartheta_{\mathrm{c}}(E)$ exponentially as $n$ increases.

We adopt the standard observer-lens-source geometry in the thin-lens approximation~\cite{Schneider:1992bmb}. The lens is represented by a thin-lens plane passing through $\mathrm{L}$ and orthogonal to the optical axis. The source $\mathrm{S}$ lies on the source plane, located at distance $D_{\mathrm{OS}}$ from $\mathrm{O}$ along the optical axis, and $D_{\mathrm{LS}}$ denotes the distance between the lens and source planes. In the asymptotically flat lens configuration considered here, we take $D_{\mathrm{OS}}=D_{\mathrm{OL}}+D_{\mathrm{LS}}$. The angle $\beta\ge 0$, measured at $\mathrm{O}$ with respect to the optical axis, denotes the source angular position.

In the small-angle and small-offset regime, $\beta, \vartheta, |\Delta \alpha_n|\ll 1$, the tangent form of the lens equation under the thin-lens approximation~\cite{Virbhadra:1999nm},
\begin{align}
\tan \beta=\tan \vartheta -\frac{D_{\mathrm{LS}}}{D_{\mathrm{OS}}} \left[\:\!
\tan \vartheta+\tan(\hat{\alpha}-\vartheta)
\:\!\right],
\label{eq:tanlenseq}
\end{align}
reduces to the strong-deflection lens equation
\begin{align}
\beta\simeq \vartheta-\frac{D_{\mathrm{LS}}}{D_{\mathrm{OS}}} \Delta \alpha_n. 
\label{eq:lenseq}
\end{align}

We now solve Eq.~\eqref{eq:lenseq} perturbatively around the zeroth-order position $\vartheta_n^0$. Let $\Delta\vartheta_n:=\vartheta-\vartheta_n^0$. Since $\varepsilon/\varepsilon_n=1+\Delta \vartheta_n/(\vartheta_{\mathrm{c}} \varepsilon_n)$, we assume $|\Delta \vartheta_n/(\vartheta_{\mathrm{c}} \varepsilon_n)|\ll 1$. Using Eqs.~\eqref{eq:angular-SDL} and~\eqref{eq:vartheta-n0-def}, we obtain
\begin{align}
\Delta \alpha_n 
\simeq -\bar{a} \log \left(
1+\frac{\Delta \vartheta_n}{\vartheta_{\mathrm{c}}\varepsilon_n}
\right)
\simeq -\bar{a}\,\frac{\Delta \vartheta_n}{\vartheta_{\mathrm{c}} \varepsilon_n}.
\label{eq:Danlin}
\end{align}
Substituting Eq.~\eqref{eq:Danlin} into Eq.~\eqref{eq:lenseq}, we have
\begin{align}
\beta\simeq \vartheta_n^0+\Delta\vartheta_n+\frac{\bar{a}D_{\mathrm{LS}}}{D_{\mathrm{OS}}} \,\frac{\Delta\vartheta_n}{\vartheta_{\mathrm{c}} \varepsilon_n}. 
\end{align}
Solving this equation for the expansion parameter, we find
\begin{align}
\frac{\Delta\vartheta_n}{\vartheta_{\mathrm{c}} \varepsilon_n}\simeq \frac{\beta-\vartheta_n^0}{\vartheta_{\mathrm{c}} \varepsilon_n+\dfrac{\bar{a}D_{\mathrm{LS}}}{D_{\mathrm{OS}}}}
\simeq \frac{ D_{\mathrm{OS}}}{\bar{a} D_{\mathrm{LS}}}(\beta-\vartheta_n^0), 
\label{eq:Dtheta}
\end{align}
where, in the last step, we have used the standard strong-lensing hierarchy holding for the astrophysical configurations considered here, 
\begin{align}
\vartheta_{\mathrm c}\varepsilon_n
\ll
\frac{\bar{a} D_{\mathrm{LS}}}{D_{\mathrm{OS}}},
\label{eq:suppress}
\end{align}
while $\bar{a} D_{\mathrm{LS}}/D_{\mathrm{OS}}$ is not anomalously small. Under this assumption, Eq.~\eqref{eq:Dtheta} shows that the logarithmic expansion is self-consistent for the nearly aligned sources satisfying
$\frac{D_{\mathrm{OS}}}{\bar aD_{\mathrm{LS}}}
\left|\beta-\vartheta_n^0
\right|\ll1$. 
Therefore, Eq.~\eqref{eq:Dtheta} gives
\begin{align}
\vartheta_n= \vartheta_n^0+\Delta \vartheta_n\simeq \vartheta_n^0+\vartheta_{\mathrm{c}} \varepsilon_n\frac{ D_{\mathrm{OS}}}{\bar{a} D_{\mathrm{LS}}} (\beta-\vartheta_n^0),
\label{eq:vtncorr}
\end{align}
where $\beta$ affects the relativistic image position only through a subleading displacement. To leading order, $\vartheta_n\simeq\vartheta_n^0=\vartheta_{\mathrm c}(1+\varepsilon_n)$, so that the image positions are mainly controlled by $\vartheta_{\mathrm{c}}$ and $\varepsilon_n$. In the perfectly aligned limit $\beta=0$, the relativistic images become Einstein rings. Denoting their angular radii by $\vartheta_n^{\mathrm{E}}$ and keeping the same subleading correction, we obtain
\begin{align}
\vartheta_n^{\mathrm{E}}\simeq \vartheta_n^0\left(1-\vartheta_{\mathrm{c}} \varepsilon_n \frac{D_{\mathrm{OS}}}{\bar{a}D_{\mathrm{LS}}}\right). 
\end{align}
At leading order, this reduces simply to $\vartheta_n^{\mathrm{E}}\simeq \vartheta_n^0$.

The relativistic images accumulate at the critical angular position.
Indeed, since $\varepsilon_n\to0$ as $n\to\infty$, Eq.~\eqref{eq:vtncorr} indicates that the sequence $\{\vartheta_n\}$ accumulates at the critical angular position
\begin{align}
\vartheta_\infty(E)=\lim_{n\to\infty}\vartheta_n(E)
=\vartheta_{\mathrm{c}}(E).
\end{align}
The images with $n\ge 2$, therefore, form a narrow bundle around the limiting angular position $\vartheta=\vartheta_\infty$. Their offsets from the accumulation point are of order $\vartheta_\infty \varepsilon_n$ and are exponentially suppressed as $n$ increases. When only the outermost relativistic image is resolved from the remaining packed images, we define the relevant angular separation as
\begin{align}
s(E)=\vartheta_1(E)-\vartheta_\infty(E).
\end{align}
Using Eqs.~\eqref{eq:thetacepn} and~\eqref{eq:epsilon-n}, and neglecting
the source-dependent subleading correction in Eq.~\eqref{eq:vtncorr}, we obtain
\begin{align}
s(E)\simeq \vartheta_\infty(E) \exp\left(\frac{\bar{b}(E)-2\pi}{\bar{a}(E)}\right).
\label{eq:sab}
\end{align}
Thus, the ratio $s(E)/\vartheta_\infty(E)$ is independent of the source
position at the leading order and directly probes the combination
$(\bar{b}-2\pi)/\bar{a}$.

We next evaluate the magnification of each relativistic image. Since the source-dependent correction of the image position is already subleading, it is sufficient at the present order to evaluate the Jacobian of the lens mapping at the zeroth-order image position $\vartheta=\vartheta_n^0$. For a point source with $\beta>0$, the magnification is
\begin{align}
\mu_n(E)=\left(\frac{\beta}{\vartheta} \frac{\partial \beta}{\partial \vartheta}\right)^{-1}\bigg|_{\vartheta=\vartheta_n^0}. 
\end{align}
From the lens Eq.~\eqref{eq:lenseq} and the linearized deflection angle~\eqref{eq:Danlin}, we have
\begin{align}
\left.
\frac{\partial \beta}{\partial \vartheta}
\right|_{\vartheta=\vartheta_n^0}=
1+\frac{\bar{a}D_{\mathrm{LS}}}{D_{\mathrm{OS}}}
\frac{1}
{\vartheta_{\mathrm{c}}\varepsilon_n}
\simeq
\frac{\bar{a}D_{\mathrm{LS}}}{D_{\mathrm{OS}}}
\frac{1}
{\vartheta_{\mathrm{c}}\varepsilon_n},
\end{align}
where the last approximation follows from Eq.~\eqref{eq:suppress}. Using $\vartheta_n^0=\vartheta_{\mathrm{c}}(1+\varepsilon_n)$ and
$\vartheta_\infty=\vartheta_{\mathrm{c}}$, we obtain
\begin{align}
\mu_n=\frac{D_{\mathrm{OS}}}{\bar{a}D_{
\mathrm{LS}}}
\frac{\vartheta_{\infty}^2}{\beta } \varepsilon_n(1+\varepsilon_n).
\end{align}
Thus, the magnifications decrease exponentially with $n$.

When the outermost relativistic image is resolved from the unresolved bundle of images with $n\ge2$, we define the corresponding flux ratio by 
\begin{align}
\mathcal{R}_{\mathrm{F}}(E)=\frac{\mu_1(E)}{\sum_{n=2}^\infty \mu_n(E)}. 
\end{align}
Assuming that $\exp(2\pi/\bar{a})\gg1$ and $\exp(\bar{b}/\bar{a})=O(1)$ in the strong-deflection regime, we obtain
\begin{align}
\mathcal{R}_{\mathrm{F}}(E)
\simeq
\exp\left(\frac{2\pi}{\bar{a}(E)}\right)=\exp\left[\:\!2\pi \kappa_{\mathrm{c}}(E)\:\!\right],
\end{align}
where the last equality uses the local relation~\eqref{eq:result}. This flux ratio is independent of the source position to leading order. Inverting these expressions, we obtain
\begin{align}
\bar{a}(E)&\simeq\frac{2\pi}{\log \mathcal{R}_{\mathrm{F}}(E)},
\label{eq:abar-from-flux}
\\
\kappa_{\mathrm{c}}(E)&\simeq \frac{1}{2\pi}\log \mathcal{R}_{\mathrm{F}}(E). 
\label{eq:kappac-from-flux}
\end{align}
The remaining SDL coefficient $\bar b$ can be reconstructed from $s(E)$. Indeed, using Eqs.~\eqref{eq:sab} and~\eqref{eq:abar-from-flux}, we find
\begin{align}
\bar{b}(E)\simeq\bar{a}(E) \log \left(
\frac{s(E)\mathcal{R}_{\mathrm{F}}(E) }{\vartheta_\infty(E)}\right). 
\label{eq:bbrecon}
\end{align}
The limiting angular position $\vartheta_\infty$ itself fixes the critical impact-parameter scale,
\begin{align}
\eta_{\mathrm{c}}(E)
\simeq
D_{\mathrm{OL}}\vartheta_\infty(E),
\label{eq:etac-from-thetainfty}
\end{align}
provided that the lens distance $D_{\mathrm{OL}}$ is known. Therefore, for each fixed specific energy $E>1$, the energy-resolved observables $\vartheta_\infty$, $s$, and $\mathcal{R}_{\mathrm{F}}$ determine the leading SDL data $\eta_{\mathrm{c}}$, $\bar a$, $\bar b$, and equivalently $\kappa_{\mathrm{c}}$. We emphasize that, while Eq.~\eqref{eq:bbrecon} reconstructs $\bar b(E)$ from observables, its theoretical prediction requires the regular part $b_{\mathrm{R}}$ of the deflection integral and hence global information about the spacetime. The above inversion applies to monoenergetic, or energy-resolved, massive-particle images. If the detected signal has a finite-energy bandwidth, the observed image is a weighted superposition of the energy-dependent image positions and magnifications, so that the energy-unresolved observables do not in general correspond to the SDL data at a single value of $E$.

If the source is variable or transient, timing provides a complementary observable. For two relativistic images whose winding numbers can be temporally identified, the winding part of the coordinate-time delay at leading order in the SDL is
\begin{align}
\Delta T_{nm}^{(\mathrm{wind})}(E)\simeq
\frac{2\pi(n-m)}{\Omega_{\mathrm{c}}(E)}
\quad (n>m),
\end{align}
up to nonwinding regular terms in the full time-delay expansion. In particular, if the nonwinding regular contribution is negligible or can be modeled, the delay between the first two relativistic images gives $\Delta T_{21}^{(\mathrm{wind})}\simeq 2\pi/\Omega_{\mathrm{c}}$, and, therefore, $\Omega_{\mathrm{c}}\simeq 2\pi/\Delta T_{21}^{(\mathrm{wind})}$. This timing identification does not require the images to be angularly resolved. Even if the images with $n\ge2$ are angularly unresolved in the bundle near $\vartheta_\infty$, their echoes may still be temporally resolved. In that case, the earliest echo from this angular bundle corresponds to the $n=2$ image at leading order. When such timing information is available, the reconstructed quantities
$\eta_{\mathrm{c}}$, $\Omega_{\mathrm{c}}$, and $E$ fix the local kinematic
data on the critical orbit. Using Eqs.~\eqref{eq:vccond}, \eqref{eq:etac},
and~\eqref{eq:Omegac}, we find
\begin{align}
v_{\mathrm{c}}^2 &= \sqrt{1-\frac{1}{E^2}}\,
\eta_{\mathrm{c}} \Omega_{\mathrm{c}},
\label{eq:vcob}
\\
R_{\mathrm{c}}&=\eta_{\mathrm{c}} \frac{\sqrt{(1-v_{\mathrm{c}}^2)(E^2-1)}}{v_{\mathrm{c}}}.
\label{eq:Rcob}
\end{align}

Combining Eq.~\eqref{eq:kappac-from-flux} with the local expression for $\kappa_{\mathrm{c}}$ in Eq.~\eqref{eq:kappa_Gbar_only}, we obtain
\begin{align}
\log \mathcal{R}_{\mathrm{F}}(E)
&\simeq
2\pi\left\{\:\!
\frac{4v_{\mathrm{c}}^2-1}{2v_{\mathrm{c}}^2+1}
-\frac{R_{\mathrm{c}}^2}{v_{\mathrm{c}}^2}\left[\:\!
\frac{1+v_{\mathrm{c}}^2}{2}\bar{G}^{\:\!\mathrm{c}}_{(0)(0)}
+\bar{G}^{\:\!\mathrm{c}}_{(3)(3)}
+\frac{(1-v_{\mathrm{c}}^2)(6v_{\mathrm{c}}^2+1)}{2(2v_{\mathrm{c}}^2+1)}\bar{G}^{\:\!\mathrm{c}}_{(1)(1)}
\:\!\right]
\:\!\right\}^{1/2}.
\label{eq:Rflux-curvature}
\end{align}
All kinematic prefactors in this expression are fixed by measured or reconstructed observables, namely $\eta_{\mathrm{c}}$, $\Omega_{\mathrm{c}}$, and $E$, through Eqs.~\eqref{eq:vcob} and~\eqref{eq:Rcob}. This form shows that the flux ratio constrains only the single local curvature combination appearing in the square brackets. In general relativity, this relation reduces to
\begin{align}
\log \mathcal{R}_{\mathrm{F}}(E)
&\simeq
2\pi
\left[\:\!
\frac{4v_{\mathrm{c}}^2-1}{2v_{\mathrm{c}}^2+1}
-\frac{4\pi R_{\mathrm{c}}^2}{v_{\mathrm{c}}^2(2v_{\mathrm{c}}^2+1)}\mathcal{S}_{\mathrm{c}}\:\!\right]^{1/2},
\label{eq:Rflux_matter}
\end{align}
where $\mathcal{S}_{\mathrm{c}}$ is given by Eq.~\eqref{eq:Sc1}. Thus, after the kinematic quantities have been reconstructed, the flux ratio constrains the single matter combination $\mathcal{S}_{\mathrm{c}}$.

In the high-energy limit $E\to \infty$, the critical timelike orbit approaches the circular photon orbit, $R_{\mathrm{c}} \to R_{\mathrm{p}}$, and the locally measured orbital speed satisfies $v_{\mathrm{c}}^2\to 1$. Equivalently, using Eq.~\eqref{eq:vcob}, we obtain $\eta_{\mathrm{c}}\Omega_{\mathrm{c}}\to 1$. Thus, the reconstruction of $R_{\mathrm{c}}$ from $\eta_{\mathrm{c}}$, $\Omega_{\mathrm{c}}$, and $E$ becomes degenerate in the strict high-energy limit. The limiting observables determine the critical impact parameter, but not $R_{\mathrm{p}}$ separately. The flux ratio $\mathcal{R}_{\mathrm{F}}^{\mathrm{p}}=\lim_{E\to \infty} \mathcal{R}_{\mathrm{F}}(E)$ is related to the local curvature as 
\begin{align}
\log \mathcal{R}_{\mathrm{F}}^{\mathrm{p}}
&\simeq
2\pi\left[\:\!
1-R_{\mathrm{p}}^2\bigl(\bar{G}^{\:\!\mathrm{p}}_{(0)(0)}+\bar{G}^{\:\!\mathrm{p}}_{(3)(3)}\bigr)
\:\!\right]^{1/2}.
\end{align}
In general relativity, this relation reduces to
\begin{align}
\log \mathcal{R}_{\mathrm{F}}^{\mathrm{p}}
&\simeq
2\pi\left[\:\!
1-8\pi R_{\mathrm{p}}^2\bigl(\rho_{\mathrm{p}}+\Pi_{\mathrm{p}}\bigr)
\:\!\right]^{1/2}.
\end{align}
Here, the label $\mathrm p$ indicates evaluation on the circular photon orbit. These high-energy relations, therefore, constrain the single local combination $R_{\mathrm{p}}^2(\bar{G}^{\mathrm{p}}_{(0)(0)}+\bar{G}^{\mathrm{p}}_{(3)(3)})$, or, in general relativity, $R_{\mathrm{p}}^2(\rho_{\mathrm{p}}+\Pi_{\mathrm{p}})$, rather than determining $R_{\mathrm{p}}$ separately.

The observables derived above are most naturally applicable to energy-resolved signals of neutral massive particles in the eikonal/geodesic regime, such as neutrinos from core-collapse supernovae, compact-object transients, or flaring active galactic nuclei, in rare source-lens-observer alignments involving Sgr A* or another supermassive black hole. Related massive-particle lensing scenarios have been discussed in Refs.~\cite{Mena:2006ym,Eiroa:2008ks,Taak:2022xdp}. The energy dependence of these observables is the characteristic signature of massive-particle lensing: finite-$E$ effects shift the accumulation angle, image separations, relative fluxes or fluences, and echo timescales relative to the photon limit. For ultrarelativistic neutrinos, these shifts are expected to be small, but the energy-resolved analysis provides a systematic way to translate either a detected deviation or a nondetection into constraints on near-critical timelike dynamics.

Recent black hole images of Sgr~A* and M87* provide photon strong-field angular benchmarks, rather than direct measurements of the massive-particle observables. For neutrino realizations, direct measurements of $\vartheta_{\infty}(E)$ and $s(E)$, together with flux or fluence ratio, would require directional reconstruction, energy resolution, and event statistics beyond current observations. A full SDL reconstruction should, therefore, be regarded as a long-term multimessenger goal. A more accessible near-term probe may be timing: delayed echoes from a sufficiently bright transient could constrain $\Omega_{\mathrm{c}}(E)$ through their inter-echo separations, while relative fluences, after modeling the source, propagation effects, and detector response, could constrain the fluence analog of $\mathcal{R}_{\mathrm{F}}(E)$ and hence $\kappa_{\mathrm{c}}(E)$. Near-term observations are thus more likely to yield upper limits or consistency tests with the null geodesic limit than a full reconstruction.

%%%%%
\section{Conclusions}
\label{sec:9}
%%%%%
We have developed a unified SDL framework for the scattering of massive particles in asymptotically flat, static, and spherically symmetric spacetimes, showing that it admits a smooth high-energy limit. For fixed $E\ge 1$, the exact critical trajectory asymptotically approaches the unstable circular orbit at $r=r_{\mathrm{c}}(E)$, while near-critical scattering trajectories execute an arbitrarily large number of windings around it before returning to infinity. The deflection angle has the universal asymptotic form $\hat{\alpha}(R_0,E)=-a\log\delta+b+O(\delta\log\delta)$, where $\delta$ measures the deviation of the turning point from the critical trajectory. For $E>1$, this may equivalently be written as $\hat{\alpha}(\eta,E)=-\bar{a}\log\varepsilon+\bar{b}+O(\varepsilon^{1/2}\log\varepsilon)$ with $\varepsilon=\eta/\eta_{\mathrm{c}}-1$. In either representation, the leading logarithmic coefficient is determined entirely by local data at the unstable circular orbit, while the constant term encodes a nonlocal contribution through the regular part of the deflection integral. For $E>1$, our impact-parameter expansion agrees exactly with the general strong-deflection formula for unbound massive particles derived in Ref.~\cite{Feleppa:2024kio}.

A central result is the covariant identification $\bar{a}=1/\kappa_{\mathrm{c}}$ for $E>1$ [Eq.~\eqref{eq:result}] or $a=2/\kappa_{\mathrm{c}}$ for $E\ge 1$ [Eq.~\eqref{eq:result2}], where $\kappa_{\mathrm{c}}$ is the radial instability exponent of the critical trajectory measured per unit azimuthal angle. We have expressed $\kappa_{\mathrm{c}}$ in terms of kinematic data on the circular orbit (e.g., $v_{\mathrm{c}}$) and curvature components, in both comoving and static frames. The coefficient $\bar{a}$ admits a coordinate-invariant characterization, independent of any particular coordinate representation of the metric. Because these relations use only the metric and geodesic motion, they apply equally to static, spherically symmetric metrics in alternative metric theories, provided that test particles follow geodesics.

As a consistency check, we have shown that our formulas recover the expected behavior in the high-energy limit. In the limit $E\to\infty$, the timelike critical trajectory approaches the corresponding null critical trajectory associated with the unstable circular photon orbit. Consequently, our expressions reduce to the standard SDL results for null geodesics.

In general relativity, the matter dependence of $\kappa_{\mathrm{c}}$ enters only through a single local scalar combination $\mathcal{S}_{\mathrm{c}}$ built from the stress-energy components measured by static observers. The sign of $\mathcal{S}_{\mathrm{c}}$ indicates whether matter tends to weaken or enhance the instability of the critical trajectory, i.e., to decrease or increase $\kappa_{\mathrm{c}}$ (and hence to increase or decrease $\bar{a}$ through $\bar{a}=1/\kappa_{\mathrm{c}}$). This provides a compact characterization of how the local energy density and (possibly anisotropic) pressures affect the strong deflection of massive particles through the stability of the critical trajectory.

We have also connected these SDL parameters with strong-lensing observables. For fixed $E>1$, the impact-parameter form of the SDL is directly tied to the angular positions and magnifications of relativistic images: the limiting angular position, the separation of the outermost relativistic image from the remaining image bunch, and the corresponding flux or fluence ratio determine $\eta_{\mathrm{c}}$, $\bar{a}$, and $\bar{b}$, and hence \(\kappa_{\mathrm{c}}\) through $\bar a=1/\kappa_{\mathrm{c}}$. Timing information from relativistic echoes provides a complementary probe of $\Omega_{\mathrm{c}}$, which, together with \(\eta_{\mathrm{c}}\) and $E$, fixes the local kinematic data on the critical orbit. These relations show how energy-resolved angular, flux, and timing measurements can in principle constrain the local curvature combination, or in general relativity, the matter combination $S_{\mathrm{c}}$, that controls the leading SDL coefficient, although a full reconstruction for massive particles should be regarded as a long-term multimessenger goal.

Finally, we comment on connections to orbit-instability diagnostics commonly used in other strong-field observables. Since $\kappa_{\mathrm{c}}$ is defined per unit azimuthal angle $\varphi$, it can be converted to the conventional instability exponent defined with respect to the Killing time $t$ using the orbital frequency $\Omega_{\mathrm{c}}=(\mathrm{d}\varphi/\mathrm{d}t)_{\mathrm{c}}$ of the unstable circular orbit, $\lambda_{\mathrm{L}}=\Omega_{\mathrm{c}} \kappa_{\mathrm{c}}$. In the high-energy limit, $\lambda_{\mathrm{L}}$ reduces to the standard Lyapunov exponent that controls, for example, the eikonal QNM damping in black holes (see, e.g., Refs.~\cite{Ferrari:1984zz,Cardoso:2008bp}). Here, however, we have obtained the corresponding instability exponent covariantly from the geodesic deviation equation and expressed it in terms of local curvature components. For finite $E$, $\lambda_{\mathrm{L}}(E)$ provides a natural time-domain measure of the instability of the critical timelike trajectory, which governs the strong deflection of massive particles and related near-separatrix phenomena.

In summary, these results provide a direct dynamical interpretation of the leading SDL coefficient as the inverse of the radial instability exponent per unit azimuthal angle along the critical trajectory. Our derivation, based on the geodesic deviation equation, further clarifies that strong deflection is governed by tidal effects---i.e., the local dynamical instability that drives geodesic deviation---in the vicinity of the unstable circular orbit approached by the critical trajectory. Moreover, because this instability exponent can be expressed in terms of local curvature quantities evaluated on the unstable circular orbit, the leading SDL coefficient is likewise determined by those local curvature data. The present analysis is restricted to asymptotically flat, static, and spherically symmetric spacetimes, and to scattering configurations defined at infinity. Natural extensions include finite-distance lensing geometries, dispersive propagation (e.g., in a plasma), time-dependent backgrounds (e.g., accreting spherically symmetric spacetimes, where dynamical photon spheres, time-evolving shadows, and QNMs have been discussed~\cite{Koga:2022dsu,Yoo:2025tzq}), and less symmetric settings, most notably stationary and axisymmetric spacetimes (see, e.g., Ref.~\cite{Igata:2025hpy} for the null case). The extension to dispersive propagation is motivated both by existing strong-deflection analyses of photon propagation in plasma~\cite{Tsupko:2013cqa,Perlick:2017fio,Bisnovatyi-Kogan:2017kii,Matsuno:2020kju} and by curvature-based formulations of lensing in terms of geometrical and matter-field data. In the weak-deflection regime, optical scalars and deflection angles have been expressed through projected Ricci and Weyl scalars and, for spherically symmetric matter distributions, through energy-momentum components~\cite{Gallo:2011mv,Crisnejo:2017jmx}. This line of work has also been extended to dispersive media, including cold plasma, and to massive charged-particle deflection using optical and Jacobi metrics~\cite{Crisnejo:2019xtp}. Adapting the present SDL analysis to such systems would require replacing the critical timelike geodesic of the physical metric by the appropriate near-critical trajectory in an effective geometry and identifying the corresponding radial instability exponent. A full treatment of wave-optical effects for massive fields in the strong-deflection regime, going beyond the present geodesic deviation-based analysis and the eikonal approximation, is left for future work.

\begin{acknowledgments}
The authors gratefully acknowledge Akihito Katsumata, Tatsuhiko Koike, Yasusada Nambu, Sousuke Noda, Kota Ogasawara, Hiromi Saida, Tetsuya Shiromizu, Mikiya Takahashi, Rohta Takahashi, Ryutaro Tomomatsu, and Hirotaka Yoshino for helpful comments and discussions. T.I. was supported in part by Gakushuin University, by JSPS KAKENHI Grants No.~JP22K03611, No.~JP23KK0048, No.~JP24H00183, and No.~JP26K07089, and by the Inamori Foundation through the Inamori Incubate Research Grants. Y.T. was supported in part by JSPS KAKENHI Grants No.~JP24K00633, No.~JP25K01034, and No.~JP26K07089. 
\end{acknowledgments}

\section*{DATA AVAILABILITY}
There are no publicly available research data or software supporting this manuscript. Requests for further information or data should be sent to the authors.

\end{document}